\documentstyle[12pt,aasms4,psfig]{article}

\newcommand\pp{\phantom*}
\newcommand\etal{{\it et al.}\ }
 
\begin{document}
 
\title{The Ultraviolet Spectroscopic Properties of Local Starbursts:
Implications at High-Redshift}
 
\author{Timothy M. Heckman$^{1,5}$, Carmelle Robert$^2$, Claus Leitherer$^3$,
Donald R. Garnett$^4$, and Fabienne van der Rydt$^3$}
 
\parindent=0em
\vspace{5cm}
 
1. Department of Physics and Astronomy, The Johns Hopkins University,
Baltimore, MD 21218; heckman@pha.jhu.edu

2. D\'epartement de Physique, Universit\'e Laval, and Observatoire du Mont 
M\'egantic, Qu\'ebec, QC G1K 7P4, Canada; carobert@phy.ulaval.ca

3. Space Telescope Science Institute, 3700 San Martin Drive,
Baltimore, MD 21218; leitherer@stsci.edu

4. Department of Astronomy, University of Minnesota, 116 Church St. SE,
Minneapolis, MN 55455; garnett@astro.spa.umn.edu

5. Adjunct Astronomer, Space Telescope Science Institute.
 
\parindent=2em
 
\begin{abstract}
We report the results of a systematic study of the vacuum-
ultraviolet ($\lambda \simeq$ 1150 to 2000 \AA) spectra of a sample
of 45 starburst and related galaxies observed with the IUE satellite. These
span broad ranges in metallicity (from 0.02 to 3 times
solar), bolometric luminosity ($\sim 10^{7}$ to $4 \times 10^{11}
L_{\odot}$), and galaxy properties (e.g. including low-mass
dwarf galaxies, normal disk galaxies, and massive galactic
mergers). The projected size of the IUE spectroscopic aperture is
typically one to several kpc and therefore usually encompasses the entire
starburst and is similar to the aperture-sizes used for spectroscopy
of high-redshift galaxies. Our principal conclusion is that local
starbursts occupy a very small fractional volume in the multi-
dimensional manifold defined by such fundamental parameters as the
extinction, metallicity, and vacuum-UV line
strengths (both stellar and interstellar) of the starburst and the
rotation speed (mass) and absolute magnitude of the starburst's
`host' galaxy. More metal-rich starbursts are redder and more
heavily extinguished in the UV, more luminous, have stronger vacuum-
UV lines, and occur in more massive and optically-brighter host
galaxies. We advocate using these local starbursts as a `training
set' to learn how to better interpret the rest-frame UV spectra of
star-forming galaxies at high-redshift, and stress that the degree
of similarity between local starbursts and high-redshift galaxies
in this multi-dimensional parameter space can already be tested
empirically. The results on local starbursts suggest that the high-
redshift `Lyman Drop-Out' galaxies are typically highly reddened
and extinguished by dust (average factor of 5 to 10 in the UV), may
have moderately high metallicities (0.1 to 1 times solar?), are
probably building galaxies with stellar surface-mass-densities
similar to present-day ellipticals, and may be suffering
substantial losses of metal-enriched gas that can `pollute' the
inter-galactic medium.
\end{abstract}

\keywords{Galaxies: Starburst, Galaxies: Evolution, Galaxies: Formation, 
Ultraviolet: Galaxies}

\section{Introduction}

Starbursts are sites of intense star-formation that occur in the
`circum-nuclear' (kpc-scale) regions of
galaxies, and dominate the integrated emission from the `host'
galaxy (cf. Leitherer et al 1991). The implied star-formation rates
are so high that the existing gas supply may sustain the starburst
for only a small fraction of a Hubble time (in agreement with
detailed models of the observed properties of starbursts, which
imply typical burst ages of-order 10$^{8}$ years). Both optical
objective prism searches and the IRAS survey have shown that
starbursts are major components of the local universe (cf. Huchra 1977;
Gallego et al et al 1995; Soifer et al 1987). Indeed, integrated over the local
universe, the total rate of (high-mass) star-formation in
circumnuclear starbursts is comparable to the rate in the disks of
spiral galaxies (Heckman 1997; Gallego et al 1995; Tresse \& Maddux 1998).
Thus, starbursts deserve to be
understood in their own right.

Starbursts are even more important when placed in the broader
context of contemporary stellar and extragalactic astrophysics. 
The cosmological relevance of starbursts has been dramatically
underscored by one of the most spectacular discoveries in years:
the existence of a population of high-redshift (z $>$ 2) star-forming
field galaxies (cf. Steidel et al 1996; Lowenthal et al 1997). The
sheer number density of these galaxies implies that they almost
certainly represent precursors of typical present-day galaxies in
an early actively-star-forming phase. This discovery therefore
moves the study of the star-forming history of the universe into
the arena of direct observations (Madau et al 1996), and gives
added impetus to the quest to understand local starbursts. 

Observations in the vacuum-UV spectral regime are crucial for both
understanding local starbursts, and for relating them to galaxies
at high-redshift. This is the spectral regime where we can most clearly
observe the direct spectroscopic signatures of the hot stars that
provide most of the bolometric luminosity of starbursts
(e.g. Sekiguchi \& Anderson 1987; Fanelli, O'Connell, \& Thuan 1988;
Leitherer, Robert,
\& Heckman 1995). Moreover,
the vacuum-UV contains a wealth of spectral features, including the
resonance transitions of most cosmically-abundant ionic species
(cf. Morton 1991).
These give UV spectroscopy a unique capability for diagnosing the
(hot) stellar population and the physical, chemical, and dynamical state of
gas in starbursts.

Since ground-based optical observations of galaxies at 
high-redshifts sample the vacuum-UV portion of their rest-frame
spectrum, we cannot understand how galaxies evolved without
documenting the vacuum-UV properties of galaxies in the present
epoch. In particular, a thorough understanding of how to exploit
the diagnostic power of the rest-frame UV spectral properties of
local starbursts will give astronomers powerful tools with which
to study star-formation and galaxy-evolution in the early universe.

Accordingly, we have undertaken an analysis of the vacuum-UV
spectroscopic properties of a sample of 45 starburst and related galaxies in
the local universe using the IUE archives. In section 2, we
describe our analysis of these data. In section 3 we use the data
to document the empirical vacuum-UV spectroscopic properties of
local starbursts and to assess the dependence of these UV
properties on crucial parameters like the dust content,
metallicity, and luminosity of the starburst, and the mass and
luminosity of the `host galaxy'. We will also describe how
published analyses of HST UV spectra allow us to better understand
the empirical results from the IUE data, and in particular to
ascertain the origin of the absorption features seen in the IUE
data (stellar vs. interstellar). In section 4, we will interpret
the results from section 3, use these results to elucidate some of
the properties of starbursts at high redshift, and point-out some
potentially far-reaching implications.

\section{IUE Spectroscopy of Starbursts}

\subsection{Motivation}

Spectra of starbursts obtained with IUE have some important
advantages relative to HST spectra: 1) At present they cover a
significantly larger sample of starbursts, and therefore better
sample the multidimensional starburst parameter space. 2) The
projected size of the $10 \times 20$ arcsec IUE aperture is
typically one to several kpc, and is therefore a good match to the
circumnuclear sizes of these starbursts. It is also (to within a factor
of a few) similar to
the projected sizes of the apertures used to obtain spectra of
high-redshift galaxies. In contrast, the typical HST apertures are
about a factor 100 smaller in projected area and often cover
selected `pieces' of the starburst (e.g. individual `super star
clusters'). However, IUE data also have some serious disadvantages:
1) The IUE resolution of 6 \AA $\sim$ 1200 km s$^{-1}$ is barely
adequate to distinguish between stellar wind lines on the one hand,
and stellar photospheric or interstellar lines on the other (see
section 3 below). 2) The signal-to-noise in a typical IUE spectrum
is modest.

Taken together, the above considerations mean that the IUE data
can document the empirical vacuum-UV spectral properties of a large
sample of local starbursts, but we need to complement these data
with HST spectra having better spectral resolution and signal-to-
noise to understand the IUE data and the starbursts themselves.
Fortunately, HST spectra of a small sample of starbursts have been
published, and we can use these results for the above purposes.

\subsection{Sample Selection}

Using the atlas of Kinney et al (1993), we have compiled a sample
of 45 starburst and star-forming galaxies that have IUE SWP spectra
with a signal-to-noise of at least 10:1 in the continuum
and for which metallicity information is available.
While our sample is by no means a rigorously defined, statistically
complete one, it does span a broad range in parameter space. More
specifically, the sample covers a range in starburst bolometric
luminosity from $\sim$ 10$^{7}$ L$_{\odot}$ to $4 \times 10^{11}$
L$_{\odot}$, starburst metallicity from 2\% solar to almost three-
times solar, and `host' galaxy velocity dispersion (or rotation
speed) as measured with the HI $\lambda$21cm line ranging from 35
km s$^{-1}$ (dwarf galaxies) to 275 km s$^{-1}$ (massive spirals
and galactic mergers). 

The sample and its basic properties are listed in Table 1, grouped in
into two sub-samples ordered by right-ascension. The first consists
of the starbursts with metallicities less than 0.4 solar, while the second
second comprises the starbursts 
with metallicity greater than 0.5 solar. The
galaxy morphological type, heliocentric radial velocity ($V_H$),
and foreground Galactic extinction ($E(B-V)_{Gal}$) were all taken
from the NASA Extragalactic Database (NED). A few values for $E(B-
V)_{Gal}$ (denoted by an asterisk) are based on the HI column
densities from the survey of Stark et al (1992), assuming that
$E(B-V) = 1$ corresponds to an HI column of 4.9 $\times 10^{21}$
cm$^{-2}$. The metallicities ($Z$) are given in terms of the oxygen
abundance by number
($Z = 12 + \log(O/H)$, where $Z_{\odot} = $8.93). Further
information about these metallicities (including references) are
given in the Table and its associated notes. With a few exceptions
among the nearest galaxies, the
distances adopted
were calculated using the standard linear
Virgocentric infall model of Schechter (1980) with parameters
$\gamma = 2$, $v_{\rm Virgo} = 976$ km s$^{-1}$, $w_\odot = 220 $
km s$^{-1}$ (Bingelli et al 1987), and $D_{\rm Virgo} = 15.9$ Mpc
(i.e. $H_{0} = 75$ km s$^{-1}$ Mpc$^{-1}$). The absolute blue
magnitude ($M_{B}$) was calculated from the apparent magnitude
(also from NED) corrected for Galactic extinction (see above). The
parameter $dV_{0}$ is the measured width of the galaxy's
HI$\lambda$21cm emission-line (Huchtmeier \& Richter 1989)
corrected for galaxy inclination using the optical ellipticity of
the galaxy. This parameter is then approximately twice the galaxy
rotation speed. It is a rough estimate only, since we have
adopted a
very simple inclination correction (assuming the galaxies to be flat
and axisymmetric) and have ignored the role of turbulence or
non-circular gas motions in broadening the HI profile.
The far-IR luminosity ($L_{IR}$) was calculated
from the `FIR' parameter (the flux between 40 and 120 microns)
published in the IRAS survey (Fullmer \& Lonsdale 1989).

\subsection{Data Analysis}

The reduction of the raw IUE data to produce the flux-calibrated
set of IUE spectra has been described in detail in Kinney et al
(1993). We have further processed these spectra as follows. The
spectra were first corrected for the effects of Galactic extinction
using the values of $E(B-V)_{Gal}$ listed in Table 1 and the
standard Milky Way UV extinction curve (Seaton 1979; Howarth 1983).
We have then used the galaxy systemic velocity given in Table 1 to
de-redshift each spectrum so that the corrected wavelength scale
is in the rest-frame of the starburst.

We have then treated the spectra as power-laws and measured the
effective spectral index $\beta$, where F$_{\lambda}~\propto~ 
\lambda^{\beta}$. The fit for $\beta$ was performed over the
spectral range between 1250 and 1850~\AA\ using the IRAF task
CURFIT. Technically, $\beta$ is an average value obtained after
fitting a second-order Legendre polynomial to a plot of
log$\lambda$ {\it vs.} log$F_{\lambda}$. To account for spectral
features, which are mostly in absorption, we performed the fit
iteratively with rejection thresholds set at -1.5$\sigma$ and
+3$\sigma$. The major uncertainties in the values of $\beta$ come
primarily from the fact that a power-law does not always provide
a good fit to the data, probably due to line-blanketing in the
spectra. Therefore, to normalize each spectrum, we fit it with a
series of four spline function segments and then divided the
spectrum by the fit. The spline fits were done iteratively with the
clipping values given above.

The strengths of the principal UV spectral features were then
measured automatically using the normalized, de-redshifted spectra
and a standard set of spectral windows. The equivalent widths of
the features were measured by integrating between two continuum
points. These windows are as follows: SiII$\lambda$1260 (1243 to
1269\AA), OI$\lambda$1302 plus SiII$\lambda$1304 (1285 to 1312\AA),
CII$\lambda$1335 (1321 to 1342\AA), SiIV$\lambda$1397 (1378 to
1406\AA), CIV$\lambda$1549 (1507 to 1553\AA). We note that the CIV
window includes the interstellar SiII$\lambda$1526 line.

These
windows are broad enough to include foreground Galactic contributions
to the interstellar absorption features. Unlike IUE, HST spectra have adequate
spectral resolutions to separate the Galactic and intrinsic
starburst absorption features. Our examination of the HST UV
spectra of 9 starburst galaxies shows that the mean equivalent
widths of the Galactic CII$\lambda$1335, SiII$\lambda$1260,
SiIV$\lambda$1397, and CIV$\lambda$1549 lines are in the range 0.6
to 0.9\AA,
with a dispersion of about a factor of two. These can be compared
to the range of values seen in the IUE starburst spectra of roughly
1 to 10 \AA (see Table 2). We conclude that the foreground
contamination of the measured starburst lines is typically only at the
level of about 20\%, and is a major contaminant ($>$ 50\%) only
in a handful of the weakest-lined starbursts.

The uncertainties in
the absolute wavelength scale of the IUE spectra are large ($\sigma
\simeq 2\AA\ \simeq 400$ km s$^{-1}$), due primarily to the
uncertain distribution of the UV light within the large IUE
aperture. Also, the non-negligible contribution by foreground
Galactic gas will have more serious (and distance-dependent) consequences
for the measured
radial velocities. Thus,
we have not tried to measure radial velocities with these data.

The IUE data on most of the individual starbursts are rather noisy.
Thus, rather than listing the equivalent widths of all the above
absorption-lines separately, we have grouped them into two
categories. The first category consists of the strongest low-ionization
features: those due to SiII$\lambda$1260, OI$\lambda$1302 plus
Si$\lambda$1304, and CII$\lambda$1335. As we will argue below,
these lines are almost entirely interstellar in origin, so we
denote the average equivalent width of these features by W$_{IS}$.
The second category consists of the two high-ionization lines
SiIV$\lambda$1397 and CIV$\lambda$1549. Again, we will argue below
that these typically have a strong contribution from stellar winds
(although interstellar absorption can also be significant, and even
dominate in some cases). We
denote the average equivalent width of these features by W$_{W}$.

%The velocity shift of each feature was defined by wavelength which
%divides the line into two pieces of equal equivalent width....
%We have therefore {\it assumed} that the mean radial velocity of
%the SiII$\lambda$1260, OI$\lambda$1302 plus Si$\lambda$1304, and 
%CII$\lambda$1335 lines (which should be dominated by interstellar
%absorption) defines the true systemic velocity of the starburst.
%We have then measured the radial velocities of the
%SiIV$\lambda$1397 and CIV$\lambda$1549 lines (which contain a
%significant contribution from stellar winds) with respect to this
%velocity. NB - CARMELLE SEEMS TO HAVE USED A WEIRD SET OF REST
%WAVELENGTHS FOR THE IS LINES. I FIND NET BLUESHIFTS FOR THE IS
%LINES THAT ARE ONLY ABOUT 200 OR 300 KM/S LESS THAN FOR CIV AND
%SiIV. ALSO, THE WINDOWS ABOVE MEAN THAT CIV WILL BE STRONGLY
%CONTAMINATED BY SiII. THUS, I RECKON WE SHOULD SKIP THE VELOCITY
%STUFF ALLTOGETHER.

These results of our analysis are listed in Table 2. We parameterize
the UV
luminosity by $\lambda$P$_{\lambda}$ at 1900\AA\
based on the flux in band between 1863 and 1963\AA\ tabulated in
Kinney et al (1993). 
Our value
has been corrected for foreground Galactic
extinction, but not for any internal extinction in the starburst.
For an {\it unreddened} starburst, the quoted
quantity represents about half the bolometric luminosity (cf.
Meurer et al 1995).
The log of the ratio of the far-IR and UV luminosities uses the
value for the far-IR luminosity from Table 1. We also list the sum
of the observed far-IR and UV luminosities. This 
approximates the {\it intrinsic} UV luminosity
(i.e. the sum of absorbed and escaping UV light - cf. Wang \&
Heckman 1996; Meurer et al 1998).

Both the ratio and sum of the far-IR and UV luminosities might be
affected by the small size of the IUE aperture relative to that
of IRAS (we may be missing a significant fraction of the UV light).
This issue has been examined empirically by Meurer et al (1998).
They have evaluated the residuals in the correlation between UV
spectral slope $\beta$ and the far-IR/UV flux ratio for these starbursts
as a function of the angular size of the starburst host galaxy
in the visible. For galaxies with optical diameters larger than about
5 arcmin they find a clear systematic excess in the ratio of
far-IR/UV flux at a given value of $\beta$ (by a typical factor of about two).
For
galaxies with smaller angular sizes, there is no significant trend with angular
size. They argue that this lack of a trend, together with the
tightness of the correlation between UV color and far-IR/UV flux ratio,
implies that the IUE aperture does not miss a significant fraction
of the UV flux for galaxies with optical angular sizes smaller
than about 5 arcmin. This illustrates that the UV emission in these
starburst galaxies must be significantly more compact than the visible
band emission from the galaxy. In our sample, only eight galaxies
(NGC 4214, NGC 4449, NGC 1097, NGC 2903, NGC 3351, NGC 4258, NGC
4321, and NGC 5236) have optical diameters greater than 5 arcmin.
We have verified that none of the conclusions listed below are changed
if these galaxies are omitted from the sample.

\section{Results}

\subsection{Overview: Diagnostics of Dust, Stars, \& Gas}

Before discussing the results of our analysis of the starbursts'
vacuum-UV spectra, it is useful to briefly summarize how such
spectra can be used to diagnose the interstellar and stellar
components of starbursts.

\subsubsection{The UV Continuum: Probe of Dust}

The effect of dust on the UV properties of starbursts is profound.
Previous papers have established that various independent
indicators of dust extinction in starbursts 
correlate strongly with one another. Calzetti et al (1994;1996)
show that the spectral slope in the vacuum-UV continuum (as
parameterized by $\beta$, where F$_{\lambda} \propto\
\lambda^{\beta}$) correlates strongly with the nebular extinction
measured in the optical using the Balmer decrement. Meurer et al
(1997) show that $\beta$ correlates well with the ratio of
far-IR to vacuum-UV flux: the greater the fraction of the UV that
is absorbed by dust and re-radiated in the far-IR, the redder the
vacuum-UV continuum. The interpretation of these correlations with
$\beta$ in terms of the effects of dust are particularly plausible
because the {\it intrinsic} value for $\beta$ in a starburst is a
relatively robust quantity. Figures 31 and 32 in Leitherer \& Heckman (1995)
show that $\beta$ should have a value between about -2.0 and -2.6
for the range of ages, initial mass functions, and ages appropriate for
starbursts.

Our detailed understanding of the above results is incomplete,
since they must involve both the geometrical distribution of the
dust, gas, and stars in the starburst and the vacuum-UV extinction
law for the dust. However, the available data strongly suggest that
(quite surprisingly) much of the dust responsible for the
extinction is apparently distributed in the
form of a moderately inhomogeneous foreground screen or `sheath'
surrounding the starburst (Gordon, Calzetti, \& Witt 1997 and
references therein). In the context of the present paper, this
result has the happy implication that (with the exception of the
most heavily dust-shrouded starbursts like M 82 and Arp 220, which
are not in our sample) we are able to directly study much of the
stellar population responsible for `powering' the starburst via
observations in the vacuum-UV (albeit often through several
magnitudes of attenuation). We will therefore adopt the quantities
$\beta$ and L$_{IR}$/L$_{UV}$ (Table 2) as indicators of dust
reddening and extinction respectively in the rest of this paper.

\subsubsection{The UV Lines: Probes of Gas \& Stars}

The vacuum-UV spectra of starburst are characterized by strong
absorption features. These absorption features can have three
different origins: stellar winds, stellar photospheres, and
interstellar gas. 

Detailed analyses of HST and HUT spectra of starbursts show that
the resonance lines due to species with low-ionization potentials
(OI, CII, SiII, FeII, AlII, etc.) are primarily interstellar in
origin. In contrast, the resonance lines due to high-ionization
species (NV, SiIV, CIV) can contain significant contributions from
both stellar winds and interstellar gas, with the relative
importance of each varying from starburst to starburst (Conti et
al 1996; Leitherer et al 1996; Heckman \& Leitherer 1997; 
Gonzalez-Delgado et al 1998a,b; Robert et al 1998). The stellar wind
contribution to these lines can also be seen in the IUE data
themselves. Robert, Leitherer, \& Heckman (1993) have studied a
composite IUE spectrum of 13 starbursts drawn from the present
sample, and conclude that a significant stellar wind contribution
to the CIV and SiIV lines is indicated by the large blueshifts of
these lines. As a counter-example, NGC 1705 is a case in which the
interstellar medium actually dominates the CIV and SiIV absorption
(York et al 1990; Heckman \& Leitherer 1997; Sahu \& Blades 1997).
The most unambiguous detection of stellar photospheric lines is
provided by excited transitions, which are usually rather weak (cf.
Heckman \& Leitherer 1997).

%As we will show in section 3.2.2 below, a significant stellar wind
%contribution to the CIV and SiIV lines in the IUE spectra is also
%indicated by the blueshifts of these lines.

As a guide to the discussion to follow, in Figure 1 we present two
high signal-to-noise vacuum-UV spectra formed by averaging together
the IUE spectra of our large sample of starburst galaxies.
Each individual normalized de-redshifted spectrum was weighted by
square of the average signal-to-noise ratio in the continuum (to
optimize the signal-to-noise in the final composite spectrum).
The first template
(`low-metallicity starburst') was formed from those objects in Table 1 with
metallicity less than 0.4 solar (mean of 0.16 solar). The
second (`high-metallicity starburst') was formed from those
objects in Table 1 with metallicity greater than 0.5 solar
(mean of 1.1 times solar). 

Figure 1 indicates the identification of the strongest absorption
and emission features, along with the likely origin of the feature
(e.g. stellar-wind, photospheric, interstellar). It is interesting
to note that the vacuum-UV spectra of starbursts are dominated by
strong {\it absorption}-lines, but the nebular {\it emission}-lines are very
weak (the reverse of the situation in the visible - cf. Leitherer,
1997).

\subsection{Metallicity Dependences}

\subsubsection{Dust}

Apart from the effects of dust, we find that the starburst's
metallicity ($Z$) is the single most important parameter in
determining its vacuum-UV properties (see Figure 1). In fact,
metallicity and the effects of dust are well-correlated.

In Figure 2a,b we show the strong correlations between $Z$ and the
dust-indicators $\beta$ and L$_{IR}$/L$_{UV}$ (see Table 3). At low
metallicity ($<$10\% solar) a significant fraction of the intrinsic
vacuum-UV actually escapes the starburst (L$_{IR}$/L$_{UV}$ $\sim$
unity), and the vacuum-UV colors are consistent with the intrinsic
(unreddened) colors expected for a starburst population ($\beta
\sim$ -2.5). In contrast, at high metallicities ($>$ solar) 90\%
to 99\% of the energy is emerging in the far-IR (L$_{IR}$/L$_{UV}$
= 10 to 100) and the vacuum-UV colors are very red ($\beta \sim$
0). Storchi-Bergmann, Calzetti, \& Kinney (1994 - hereafter SBCK)
had previously noted the correlation between $Z$ and UV color.

We believe these correlations have a straightforward
interpretation: the vacuum-UV radiation escaping from starbursts
suffers an increasing amount of reddening and extinction as the
dust-to-gas ratio in the starburst ISM increases with metallicity.
This will be true provided that neither the gas column density
towards the starburst, nor the fraction of interstellar metals
locked into dust grains are strong functions of $Z$.

\subsubsection{The Stellar Population}

The properties of the vacuum-UV stellar absorption-lines are also
strongly dependent on metallicity. In Figure 3 we show the
correlation of the strengths of the high-ionization (CIV and SiIV)
lines with $Z$. Note that the CIV line has a weak stellar emission
component, and that we have measured the net absorption strength.
This $Z$-dependence
(noted previously by SBCK) is not surprising, given the likely
strong contribution to these lines by from stellar winds. 
Theoretically, we expect that since hot-star winds are
radiatively driven (Castor, Abbott, \& Klein 1975),
the strengths of the vacuum-UV stellar wind
lines will be metallicity-dependent. This is confirmed by available
HST and HUT spectra of LMC and especially SMC stars (Walborn et al
1995; Puls et al 1996). If post-main-sequence stars (supergiants
and Wolf-Rayet stars) contribute significantly to the integrated
light, the stellar wind properties enter in a second, indirect way.
That is, the spectral synthesis models heavily depend on the
evolutionary history of the OB population, which in turn is
critically dependent on the stellar mass-loss rates, which in turn
are a function of the metal abundance (see Maeder \& Conti 1994).

The high signal-to-noise ratios in the composite spectra shown in
Figure 1 allow us to look for the generally-weak absorption-lines
that are of stellar-photospheric origin (because they correspond
to transitions out of highly excited states). In the
high-metallicity composite spectrum we identify the following 
lines of predominantly stellar
photospheric origin: CIII$\lambda$1175, the blend of
SiIII$\lambda$1417 and CIII$\lambda\lambda$1426,1428,
SV$\lambda$1502, SiIII$\lambda$1892, and
FeIII$\lambda\lambda$1925,1960. These features are significantly
weaker (see Figure 1) in the low-metallicity composite spectrum,
and we can only tentatively detect a few lines (possibly
CIII$\lambda$1892, FeIII$\lambda$1925, and FeIII$\lambda$1960).
In HST spectra of starbursts, the blend of SiIII photospheric lines
centered near 1299 \AA\ is relatively strong. With the low spectral
resolution of IUE this feature is inextricably blended with the
strong SiII+OI interstellar lines at $\sim$ 1303 \AA.

\subsubsection{The Gas}

Finally, we find that the properties of both the interstellar
absorption-lines and nebular emission-lines also depend strongly
on starburst metallicity. As shown in Figure 4, the strengths of
the low-ionization absorption-lines (which HST and HUT spectra show
to be primarily interstellar in origin) correlate well with
metallicity, but with a very shallow slope. This is not too
surprising. Analyses of HST and HUT spectra (cf. Heckman \&
Leitherer 1997; Sahu \& Blades 1997; Gonzalez-Delgado et al
1998a,b; Pettini \& Lipman 1995) show that the strong interstellar
lines are saturated (highly optically-thick). In this case, the
line equivalent width is only weakly dependent on the ionic column
density, and more strongly on the velocity dispersion of the
absorbing gas: $W \propto b[ln(N_{ion}/b)]^{0.5}$, where $b$ is
the normal Doppler line-broadening parameter. Over the range that
we sample well, the starburst metallicity increases by a factor of
almost 40 (from 0.08 to 3 solar). Accounting for the mean
contribution to the measured lines by Galactic foreground gas
(roughly 0.8\AA\, as discussed above),
the equivalent widths of
the strong interstellar lines only increase by an average factor
of 2 to 3 over this metallicity range.
This is consistent with the strong interstellar lines
being very optically-thick, at least in the high-metallicity
starbursts. Our inspection of HST spectra of starbursts implies that
the correlation in Fig. 4 also
reflects a trend for larger velocity dispersions in the absorbing
gas in the starbursts with higher metallicity.

%If we hypothesize that ionic column density is proportional to
%metallicity and that the velocity dispersion is independent of
%metallicity, then this would imply an increase in the optical
%depth at line-center from a value of about 2 up to 80 as the
%metallicity increases by a factor of 40 (cf. Table 3.1 in Spitzer
%1978). 

The vacuum-UV nebular emission-lines in starbursts are weak (and
the potentially-strong Ly$\alpha$ emission-line is corrupted by the
strong Geocoronal feature). Nevertheless, Figure 1 shows that there
is a trend for stronger nebular emission-lines (on-average) in the
starbursts of {\it lower} metallicity. This effect is seen in both
the CIII]$\lambda$1909 and CII]$\lambda$2326 (not shown in Fig. 1) 
emission-lines, so it
cannot be due to a dependence of ionization-state on $Z$.
Moreover, the fact that these are both semi-forbidden-lines implies
that the weaker line-emission at high-metallicity is not due to
preferential dust-extinction of the emission-line photons via
multiple scattering of resonance-lines (as may occur in the case
of Ly$\alpha$ - cf. Chen \& Neufeld 1994; Giavalisco et al 1996).
Instead, the correlation may be due either to a decreasing ratio
of ionizing radiation with respect to non-ionizing vacuum-UV
radiation at high-metallicity, or (more likely) to a decrease in
the equilibrium electron temperature in higher-metallicity gas that
shifts nebular cooling via collisionally-excited line-emission from
the UV to the IR.

\subsection{Correlations Between the Absorption-Line and Continuum
Properties}

In this section, we document the empirical relationships between
the dust-sensitive properties of the vacuum-UV continuum and the
other major vacuum-UV `observables', namely the strengths of the
prominent absorption features.

In Figure 5a,b we show how the strengths of the
low-ionization (CII, OI, and SiII) and high-ionization
(CIV and SiIV) resonance
absorption-lines depend upon the vacuum-UV spectral slope $\beta$.
Figure 6a,b similarly plots the strengths of the low-ionization
and high-ionization absorption-lines versus the ratio of the IR and
UV luminosities. In all four plots we see strong correlations (see
Table 3), in the sense that the absorption-lines are stronger in
starbursts whose vacuum-UV continuum suffers more severe reddening
and extinction due to dust. These correlations are somewhat
stronger for the low-ionization lines (r = 0.84 and 0.75 for Fig.
5a and 6a) than for the high-ionization lines (r = 0.76 and 0.66
for Fig. 5b and 6b). 

The stronger {\it interstellar} absorption-lines in systems having
a higher amount of dust extinction and reddening is physically
reasonable, since both trace the effect of interstellar metals on
the escaping UV light. However, as noted above, these interstellar
lines are optically-thick, and so their strength is only weakly
dependent on ionic column density. This contrasts with $\beta$,
since in a simple foreground-screen interpretation of the
reddening, the difference between the observed and intrinsic value
of $\beta$ ($\Delta$$\beta$ = $\beta_{obs}$ - $\beta_{0}$) is just
linearly proportional to the foreground dust column density. Since
we would expect the column densities of the gas-phase metal-ions
and the dust to be roughly proportional to one another, it is then
difficult to understand the strong and roughly linear dependence
of the equivalent widths of the strongly saturated interstellar
lines on $\beta$. We infer then that the strong correlation in
Figure 5a implies that there must also be a trend for the velocity
dispersion in the starburst ISM to be larger in the cases where the
effective reddening suffered by the vacuum-UV continuum is larger.
This trend is present in the small sample of starbursts
with HST UV spectra having adequate spectral resolution to measure
the line widths.

Finally, in Figure 7 we show that there is a strong positive
correlation (r = 0.69) between the strengths of the low-ionization
(mostly interstellar) lines and the high-ionization (stellar-plus-
interstellar) lines. This correlation probably reflects both the
interstellar contribution to the high-ionization lines and the
mutual dependence of the strengths of both the high- and low-
ionization lines on the metallicity (Figures 3 and 4). 

\subsection{Dependences on Starburst Luminosity}

In this section we will approximate the {\it intrinsic} starburst
UV luminosity as the sum of the {\it observed} vacuum-
UV and far-IR luminosities (roughly representing the escaping and
reprocessed UV light respectively). Except in the starbursts of the
lowest metallicity, the starburst luminosity is dominated by the
far-IR emission (see section 3.2 above).

We preface this discussion by noting that there is a potentially
significant selection effect that can affect our conclusions. Even
though our sample of starbursts is not rigorously defined, a
starburst is loosely defined to be an event that contributes
significantly to (or even dominates) the bolometric luminosity 
of its `host' galaxy. This means that our sample will be selectively
missing low-luminosity regions of star-formation in luminous galaxies.
Given the strong correlation between metallicity and galaxy absolute
magnitude (Zaritzky, Kennicutt, \& Huchra 1994), this also means that we
will be selectively missing low-luminosity metal-rich `starbursts'
(e.g. giant HII regions in luminous, metal-rich galaxies).

This effect can clearly be seen in Figure 8, which shows a highly
significant positive correlation (r = 0.59) between the luminosity
of the starburst and its metallicity. However, this correlation
cannot be entirely due to selection effects. The above effect
explains why the lower right part of the figure (high metallicity,
low luminosity) is un-populated. However, it does {\it not} explain
the scarcity of points in the upper left (low metallicity, high
luminosity). It appears that the maximum starburst luminosity scales
with metallicity. This is likely to be a secondary correlation, due to
the mutual dependence of both starburst luminosity and metallicity
on the absolute magnitude and mass of the starburst host 
galaxy (see section 3.5
below).

We also find (Figure 9a,b) that the most luminous
starbursts suffer the most severe extinction and reddening due to
dust. Only starbursts with $L_{UV}+L_{IR}<$ few $\times 10^{9}$
L$_{\odot}$ have colors expected for an unreddened starburst and
have vacuum-UV luminosities that rival their far-IR luminosities.
Starbursts that lie at or above the `knee' in the local starburst
luminosity function ($L_{UV}+L_{IR}>$ few $\times 10^{10}$ L$_{\odot}$
- cf. Soifer et al 1987) have red UV continua ($\beta \sim$ -1 to
+0.4) and are dominated by far-IR emission ($L_{IR} \sim 10 - 100 \times
L_{UV}$). The caveat discussed above is relevant: while the paucity
of luminous unreddened/unextinguished starbursts is real, the lack of
heavily reddened/extinguished low-luminosity `starbursts' in our sample
is a selection effect.

It is also worth emphasizing that the trend seen in Figure 9b is not
unique to starbursts.
Wang \& Heckman (1996) have previously shown that
normal star-forming disk galaxies (non-starburst systems) also
exhibit a strong positive correlation between UV dust opacity (the
global ratio of far-IR and UV flux) and intrinsic global UV luminosity
(the sum of the observed UV and far-IR luminosities). This is not the
result of the kind of selection effect discussed above.

Finally, in Figure 10a,b we show the correlations between starburst
luminosity $L_{UV}+L_{IR}$ and the strengths of the high-ionization and
low-ionization absorption-lines. These correlations are
significant, but weaker (r = 0.54; r = 0.46) than the correlations
of $Z$ with both $L_{UV}+L_{IR}$ (r = 0.59) and with line strengths (r
= 0.85 and r = 0.66 for the high- and low-ionization lines
respectively). Thus, the correlations shown in Figure 10 may be
secondary ones. The selection effect described above will certainly be
pertinent to Figure 10a (that is, the lack of `starbursts' with
strong stellar wind lines and low luminosity is a selection effect).
The situation regarding the interstellar lines (Figure 10b) is less
clear in this regard, since the line strengths are probably set
largely by the velocity dispersion in the gas rather than by its
metallicity.

It is interesting that none of the above parameters ($\beta$,
$L_{IR}/L_{UV}$, $Z$, or line strength) correlate with the
luminosity of the {\it escaping} UV radiation ($L_{UV}$). For
example, the correlation between the {\it intrinsic} starburst
luminosity and line strength (Figure 10) is `undone' by the inverse
correlation between intrinsic starburst luminosity and the fraction
that escapes in the UV (Figure 9b).
    
\subsection{Host Galaxy Properties \& Starburst Gas Dynamics}

Having explored the correlations among the various starburst
properties, we now briefly consider the relationship between the
starburst and its `host galaxy'. We characterize the host galaxy
in two ways: its rotation speed ($v_{rot} \simeq 1/2 dV_{0}$) and its
absolute blue
magnitude ($M_{B}$) - see Table 1. 

It is important to note that
while the quantity $v_{rot}$ should be a characteristic of the host
galaxy that is unaffected by the starburst, the starburst may in
principle make
a significant contribution to $M_{B}$. To assess this quantitatively,
we have used the sum $L_{UV}+L_{IR}$ to approximate the intinsic UV
luminosity of the starburst, and used the models of Leitherer \&
Heckman (1995) to then predict the starburst absolute blue magnitude
({\it in the absence of dust extinction}). We then find that the 
median value for the contribution of the unextinguished starburst to
the total observed blue
light from the galaxy is only about 20\%. Extinction will significantly
diminish this contribution.  
The typical values for
$L_{IR}/L_{UV}$ (e.g. Fig. 2b) together with the starburst effective
dust `obscuration law'
determined by Calzetti (1997) imply that the typical extinction
in the B-band will be about a magnitude. Thus, we conclude that
(with the possible exceptions of the few most powerful starbursts
in our sample),
the quantity $M_{B}$ primarily measures the properties of the
starburst's host galaxy.

Figures 11 and 12 show montages of the correlations of $dV_{0}$
and $M_{B}$ respectively with the principal properties of the
starburst. From this it is clear that brighter and more massive
galaxies host more-metal-rich, dustier, and more-luminous (but
note the caveat in section 3.4 above)
starbursts with stronger UV absorption-lines. In a similar vein,
Wang \& Heckman (1996) have shown that normal star-forming disk
galaxies also exhibit a strong dependence of the UV opacity (as
measured by the ratio of far-IR and UV flux) on both $M_{B}$ and
$dV_{0}$.

Apart from the very strong correlation of $M_{B}$ with $L_{UV}+L_{IR}$
(which at least in part is due to the selection effect described
in section 3.4),
the strongest correlations of both $dV_{0}$ and $M_{B}$ are with
starburst metallicity (Table 3). It has been previously shown that
the mean metallicity of the gas and stars in normal galaxies
correlates strongly and positively with the galaxy rotation speed,
stellar velocity dispersion, and absolute magnitude (cf. Bender et
al 1993; Zaritsky et al 1994; Skillman et al 1989). Thus, the
strong analogous correlations exhibited by the starbursts are
perhaps not surprising. Since we have previously shown (section
3.2) that higher-metallicity starbursts are dustier and have
stronger UV absorption-lines, the corresponding correlations with
the host galaxy properties are probably secondary in nature. Even though
there is a weak correlation between the host galaxy rotation speed
and starburst luminosity, even relatively low-mass galaxies can 
host moderately powerful starbursts like M 82 (see Lehnert \&
Heckman 1996a).

The relative weakness of the correlation between the strength of
the interstellar absorption-lines and $dV_{0}$ (Figure 11), is
especially interesting. As emphasized above, the optically-thick
nature of the strong interstellar lines means that their strengths
are (to-first-order) determined primarily by the velocity
dispersion in the ISM of the starburst. The weakness of the
correlation of line strength with $dV_{0}$ then suggests that the
dynamics of the absorbing gas are not {\it dominated} by the
gravitational potential of the host galaxy.

This inference is further supported by the enormous equivalent
widths of the starburst interstellar lines, since these require
correspondingly large velocity dispersions. For example, for gas
with a Maxwellian velocity distribution and a line-center optical
depth of 10 (100) in a strong UV resonance line, the observed
equivalent widths of 4 to 6 \AA\ in the strong-lined starbursts
imply three-dimensional velocity dispersions in the gas of
$\sigma_{3D} = 370 - 560 (270 - 400) f^{-1}$ km s$^{-1}$ 
(where $f$ is the fraction of the UV continuum that is covered by
absorbing material). These velocity dispersions are significantly
greater than the observed rotation speeds (typically 100 to 200 km
s$^{-1}$).

Of course, the most direct evidence for a non-gravitational origin
of the gas-dynamics comes from analyses of HST and HUT spectra of
individual starbursts. These show that the interstellar absorption-lines
are often blueshifted by one-to-several-hundred km s$^{-1}$
with respect to the systemic velocity of the galaxy (Heckman \&
Leitherer 1997; Gonzalez-Delgado et al 1998a,b; Lequeux et al 1995;
Kunth et al 1998). Thus, in many cases, the
absorbing gas is flowing outward from the starburst, probably
helping to `feed' the superwinds that have been studied heretofore mostly
via their optical (e.g. Lehnert \& Heckman 1996b) and X-ray (e.g.
Dahlem, Weaver, \& Heckman 1998) emission. Thus, the kinematic
properties of the strong interstellar absorption-lines in
starbursts reflect in part the hydrodynamical consequences of
the starburst, and do not straightforwardly probe the gravitational
potential of the host galaxy.

%NOTE - FOR TAU = 10 (100) A GIVEN INTRINSIC LINE-BROADENING DUE
%TO TURBULENCE (DOPPLER FWHM) WOULD YIELD A LINE EQW IS 1.82 (2.56)
%TIMES THE TRUE DOPPLER FWHM. THUS OUR MEASURED WIDTHS OF 400 TO
%1200 KM/S IMPLY INTRINSIC DOPPLER FWHM OF `ONLY' 200 TO 600 KM/S.

\section{Summary \& Implications}

The principal conclusion of this paper is that local starbursts
occupy a very small fractional volume in the multi-dimensional
manifold defined by such fundamental parameters as the
extinction, metallicity, and vacuum-UV line strengths
(both stellar and interstellar) of the starburst and the rotation
speed (mass) and absolute magnitude of the starburst's `host'
galaxy. More metal-rich starbursts are redder, more heavily
extinguished (IR-dominated), have stronger vacuum-UV
lines, occur in more massive and brighter host galaxies, and
can be more luminous.
 
Starbursts with solar or higher metallicity are heavily extinguished
indeed: only 1 to 10\% of the vacuum-UV radiation escapes and the resulting
UV spectral energy distribution is nearly flat in F$_{\lambda}$ (Fig.
2). Only the most metal-poor starbursts ($<$ 10\% solar) are
relatively unaffected by dust (Fig. 2). These correlations are not
surprising, since we expect the dust-to-gas ratio in the
starburst's ISM to scale roughly with metallicity. 

In principle,
relatively red vacuum-UV spectral energy distributions could be due
to the effects of the age of the stellar population or the form of
the initial mass function (IMF)
rather than dust. That is, the redder spectra
might correspond to older starbursts, starbursts whose UV spectrum
was more heavily contaminated by an older underlying stellar population,
or starbursts with a steeper (less O-star-enriched) IMF.
However, the strong correlation between the UV color ($\beta$) and
the strengths of the CIV and SiIV (predominantly) stellar wind lines 
(Fig. 5b) shows
that this cannot be the case: an age or IMF effect would require the stellar
wind features due to O stars to be {\it weaker} in the redder
starbursts - exactly the opposite of what we observe.
 
The metal-poor starbursts have systematically weaker vacuum-UV
absorption-lines (Fig. 1). In the case of the stellar wind lines
(Fig. 3), this result agrees with expectations based on both theory
and observations of metal-poor, hot, high-mass stars. In the case
of the interstellar lines (Figure 4), this probably reflects both
higher average ionic column densities and larger velocity
dispersions in the more metal-rich starbursts. That is, even though
the strong interstellar lines are optically-thick (and their strength
will thus be set to first-order by the velocity dispersion), 
there will still
be a weak dependence of equivalent width on column density.
Moreover, the more metal-rich starbursts can also be more powerful (Fig. 8) and
are situated in more massive galaxies (Fig. 11). We expect them to have
higher average ISM velocity dispersions, since both gravity and the
mechanical `stirring' provided by supernovae and stellar winds will
contribute to the gas dynamics.
 
The trend for the more metal-rich starbursts to occur in brighter
and more massive host galaxies (Fig. 11 and 12) presumably reflects the
well-known mass-metallicity relation defined by normal galaxies
(e.g. Zaritsky, Kennicutt, \& Huchra 1994). That is, the ISM in
massive galaxies will already be pre-enriched to relatively high
metallicity before the starburst is even initiated. During the
starburst itself, the fraction of the newly-synthesized metals that
are retained (rather than blown out) should also be a strong
function of the depth of the gravitational potential well (e.g.
Dekel \& Silk 1986). The correlation between the metallicity and
the luminosity of the starburst (even though it is due in part to
selection effects - see section 3.4) may
reflect other fundamental trends: more massive galaxies can host more
powerful
starbursts (Fig. 11) and have more metal-rich interstellar gas with
which to fuel the starburst. The luminosity {\it vs.} rotation-speed
connection in starbursts has been previously discussed by Heckman
(1993), Lehnert \& Heckman (1996b), and Meurer et al (1997): basic
considerations of causality in a self-gravitating system demand
that the maximum star-formation rate will scale as the cube of the
rotation speed.

The above results are not only very illuminating regarding the nature
of the starburst phenomenon, they also
have a variety of interesting implications for the interpretation
of the rest-frame-UV properties of galaxies at high-redshift.

First and foremost, starbursts in the present universe
emit most of their light in the far-infrared, not in the
ultraviolet (Fig. 9). Thus, an ultraviolet census of the local universe
would significantly underestimate the true high-mass star-formation-rate and
would systematically under-represent the most powerful, most metal-
rich starbursts (Fig. 2), occuring in the most massive galaxies (Fig. 11).
This {\it
may} also be true at high-redshift, where the current estimates of
star-formation rely almost exclusively on data pertaining to the
rest-frame vacuum-UV (e.g. Madau et al 1996, 1998). If there {\it
is} a dependence of dust extinction on luminosity at high-redshift,
this will affect the apparent shape of the luminosity function. For
example, current samples at high-redshift might under-represent
young/forming massive elliptical galaxies. These would have high
metallicities (and thus a high dust-to-gas ratio) coupled with very
large average gas surface mass densities ($\sim 10^{22} - 10^{23}$ cm$^{-2}$),
and would hence have large dust
opacities.

Using the strong correlation between the vacuum-UV color of local
starbursts ($\beta$) and the ratio of far-IR to vacuum-UV light
emitted by local starbursts, Meurer et al (1998) estimate that an
average vacuum-UV-selected galaxy at high-redshift (e.g. Steidel
et al 1996; Lowenthal et al 1997) suffers about 2 to 3 magnitudes
of UV extinction (in agreement with modeling of the rest-frame
UV-through-visible spectral energy distributions by Sawicki \& Yee
1998). The `correct' prescription for de-extincting the high-
redshift galaxies in order to correctly obtain the bolometric
luminosity and star-formation rate is a matter of on-going debate
(see Pettini 1997 and Madau et al 1996, 1998 for other viewpoints).

For example, one uncertainty at high-redshift is the unknown
relative importance of the age or the IMF of the stellar population
{\it vs.} dust
in producing the observed UV spectral energy distribution.
As we have argued above, the trend for the {\it redder} starbursts to have
{\it stronger} absorption features due to massive stars (Fig. 5b) rules
out age or the IMF as the primary determinant of UV color in our sample.
Thus, it will be extremely interesting to see if the same
correlation holds in the high-redshift galaxies.
In any case, it seems fair to conclude that the history of high-mass
star-formation in the universe at early times will remain
uncertain until the effects of dust extinction are better
understood.

The extinction corrections advocated by Meurer et al (1997) imply
large bolometric luminosities
for the high-redshift galaxies ($\sim$ 10$^{11}$ to 10$^{13}$
L$_{\odot}$ for H$_{0}$ = 75 km s$^{-1}$ Mpc$^{-1}$ and q$_{0}$ =
0.1). Interestingly, the bolometric surface-brightnesses of the
extinction-corrected high-redshift galaxies are then very similar
to the values seen in local starbursts: $\sim$ 10$^{10}$ to
10$^{11}$ L$_{\odot}$ kpc$^{-2}$ (Meurer et al 1997). The
high-redshift galaxies thus appear to be `scaled-up' (larger and more
luminous) versions of the local starbursts. The physics behind this
`characteristic' surface-brightness is unclear (cf. Meurer et al
1997; Lehnert \& Heckman 1996a). However, it is intruiging that the
implied average surface-mass-density of the stars within the half-
light radius ($\sim$ 10$^{2}$ to 10$^{3}$ M$_{\odot}$ pc$^{-2}$)
is quite similar to the values in present-day elliptical galaxies.
Are we witnessing the formation of elliptical galaxies and/or
bulges, or only the formation of precursor fragments thereof (e.g.
Sawicki \& Yee 1998; Trager et al 1997)?

If the high-redshift galaxies are indeed scaled-up versions of
local starbursts, we might expect that they would follow the same
trends summarized above. For example, do the high-
redshift galaxies should show the same strong correlation between
the strength of the UV absorption-lines (stellar and interstellar)
and $\beta$ (Fig. 5), whereby the more metal-rich local starbursts are both
redder and stronger-lined? As more near-IR (rest-frame
visible) data become available, will
the high-redshift galaxies obey the starburst correlations (Fig. 11 and 12) of
both vacuum-UV color and absorption-lines strengths with M$_{B}$ and
rotation speed (do they follow a mass-metallicity relation?). When the
high-redshift galaxies are extinction-corrected, are the more
luminous systems redder and stronger-lined in
the UV (as in the case of the local starbursts - Fig. 9 and 10 respectively)?
Finally, do the strengths of the low-ionization (primarily interstellar)
and high-ionization (significant stellar wind contribution) absorption-lines
correlate well with one-another, as in the local starbursts (Fig. 7)?

If the answers to these questions are `yes' (that is, if
the high-redshift galaxies and local starbursts do occupy the
same parts of the multi-dimensional parameter space summarized
above), we might be able to use the UV properties of the high-
redshift galaxies to `guesstimate' their metallicities. For
example, the strong correlation between vacuum-UV color ($\beta$)
and metallicity in local starbursts (Fig. 2) - if applied naively to high-
redshift galaxies - would suggest a broad range in metallicity from
substantially subsolar to solar or higher and a median value of
perhaps 0.3 solar. This is somewhat higher than the mean
metallicity in the damped Ly$\alpha$ systems (Pettini et al 1997;
Vladilo 1997; Lu et al 1997), but this may be due to selection
effects: the UV-selected galaxies are the most actively star-forming
regions of galaxies, while the damped Ly$\alpha$ systems
tend to sample the outer, less-chemically-enriched parts of
galaxies (e.g. Prochaska \& Wolfe 1997) or perhaps proto-galactic
fragments or dwarf galaxies (e.g. Vladilo 1997). Our `guesstimated'
metallicities of the high-redshift galaxies are much higher than
those suggested by Trager et al (1997), who argue that we are
witnessing the formation of Population II spheroidal systems.

It is also important to sound a cautionary note: most of the strong
absorption-lines in the spectra of local starbursts are of {\it
interstellar} rather than {\it stellar} origin. This is probably
generically true for the low-ionization resonance lines (although B stars
may make a non-negligible contribution).
Even the high-ionization resonance lines like CIV and SiIV can contain
strong interstellar contributions. In fact, in extreme cases (e.g. NGC
1705) even these lines are dominated by interstellar gas. This
highlights the difficulty in using the UV spectra of galaxies to
deduce their stellar content: data of relatively high spectral
resolution and signal-to-noise are needed to reliably isolate the
stellar and interstellar components. Simply measuring the
equivalent widths of the lines is not enough: it is the {\it
profile shapes} that contain the key information. The stellar
photospheric absorption-lines identified in Fig. 1 may be the
most straightforward features for the detection of stars in 
high-redshift galaxies, since they are not resonance lines, and so are
uncontaminated by interstellar absorption. Unfortunately, most of
these lines are weak. The relatively strong CIII$\lambda$1175 line
is lost in the Ly$\alpha$ forest at high-redshift, and the
CIII$\lambda$1892, FeIII$\lambda$1925, and FeIII$\lambda$1960 lines
are often `clobbered' by the OH night-sky lines in galaxies
at z $>$ 2.6.

Finally - based on local starbursts - it seems likely that the gas
kinematics that are measured in the high-redshift galaxies using
the interstellar absorption-lines are telling us a great deal about
the hydrodynamical consequences of high-mass star-formation on the
interstellar medium, and thus it will be difficult to use them to
straightforwardly glean information about
the gravitational potential or mass of the galaxy. Even the widths
of the optical nebular emission-lines in local starbursts are not
always reliable tracers of the galaxy potential well (cf. Lehnert
\& Heckman 1996a; but see Melnick, Terlevich, \& Moles 1988). This means that
it may be tricky to determine
masses for the high-redshift galaxies without measuring real
rotation-curves via spatially-resolved spectroscopy.

On the brighter side, there is now rather direct observational
evidence that the kinematics of the interstellar absorption-lines
in the high-redshift galaxies reflect an outflowing metal-enriched
gas, thereby allowing us to directly study the `pollution' of the
inter-galactic medium with metals in the early universe. The
signature of these outflows is interstellar absorption lines that
are systematically blueshifted by several hundred km s$^{-1}$
relative to the Ly$\alpha$ emission-line (Franx et al 1997; Heckman
1997; Pettini 1997). In local starbursts, this is due to outflowing
gas that both produces the blue-shifted absorption-lines and
absorbs-away the blue side of the Ly$\alpha$ emission-line (Lequeux
et al 1995; Gonzalez-Delgado et al 1998a,b; Kunth et al 1998). The
case for an outflow in the high-redshift galaxies could be
strengthened by using the weak stellar photospheric lines (Fig. 1)
to unambiguously determine the galaxy systemic
velocity - cf. Heckman \& Leitherer (1997) and Sahu \& Blades
(1997).

If these outflows escape the galactic potential well, we should see
their cumulative effect in the form of a metal-enriched
inter-galactic medium (IGM). The most dramatic such evidence is the
existence of an IGM in rich clusters of galaxies whose metal
content probably exceeds that of the cluster galaxies themselves.
Recent ASCA X-ray spectra of hot gas in clusters of galaxies show
approximately solar abundance ratios for the $\alpha$-process
elements like O, Ne, and Si relative to Fe (Mushotzky et al 1996;
Ishimaru \& Arimoto 1997). This implicates `core-collapse'
supernovae (the end product of high-mass stars) and - by inference
- starburst-driven superwinds as the source of at least 90\% of the
metals in the cluster IGM (Renzini 1997; Gibson, Loewenstein, \&
Mushotzky 1997). More generally, the presence of metals in the
`Ly$\alpha$ forest' at high-redshift (the cloudy component of the
early IGM) is certainly suggestive of the dispersal of chemically-
enriched material by early superwinds (cf. Cowie et al 1995; Madau
\& Shull 1996). The Ly$\alpha$ forest clouds appear to have a high
ratio of Si/C, suggestive of core-collapse supernovae as the source
(Cowie et al 1995; Giroux \& Shull 1997).

In conclusion, vacuum-ultraviolet observations give us a rich array
of diagnostic probes of the stellar and interstellar components of
starbursts in the local universe. The analysis described in this
paper strongly suggests that these starbursts obey well-defined
relationships between such key physical parameters as extinction,
metallicity, starburst luminosity and gas dynamics, and host galaxy
mass. These objects can therefore serve as local laboratories in
which we can study the processes that were important in the galaxy-
building events that can now be directly observed at high-redshift.
The results of this synergistic study of the most actively-star-
forming local and distant galaxies suggests that the latter have
rather high UV extinctions (average factor 5 to 10), may have
moderately high metallicities (typically 0.1 to 1 times solar), are
probably building galaxies having surface mass-densities comparable
to present-day ellipticals, and are suffering substantial losses
of metal-enriched outflowing gas that can explain the observed
`pollution' of the inter-galactic and intra-cluster medium.

{\bf Acknowledgments}

We would like to thank Anne Kinney and Ralph Bohlin for helping us
access and understand the IUE spectra that form the basis of this paper.
We also thank Gerhardt Meurer and
Daniela Calzetti for many enlightening discussions about the properties
of local starbursts, and ditto Mark Dickinson regarding the high-redshift
galaxies. This research was primarily supported by NASA LTSA grant NAGW-3138.
DG acknowledges the support of the NASA LTSA grant NAG5-6416. This
research has made use of the NASA/IPAC extragalactic
database (NED), which is operated by the Jet Propulsion Laboratory, Caltech,
under contract with the National Aeronautics and Space Administration.

\newpage
% Figure Captions

\figcaption{IUE spectral templates of low-metallicity (top) and high-
metallicity (bottom) starbursts, formed from the spectra in our
sample (Table 1). Each spectrum is the weighted average of the normalized
spectra of about
20 starbursts, with mean metallicities of 0.16 solar (top) and 1.1
solar (bottom). See text for details. The normalized high-metallicity
spectrum has been offset by -0.5 normalized flux units for clarity.
A number of spectral features
are indicated by tick marks and have the following identifications
(from left-to-right): CIII$\lambda$1175 (P), NV$\lambda$1240 (W),
SiII$\lambda$1260 (I), OI$\lambda$1302 plus SiII$\lambda$1304 (I)
- possibly blended with SiIII$\lambda$1299 (P),
CII$\lambda$1335 (I), SiIV$\lambda$1400 (W plus I),
SiIII$\lambda$1417 plus CIII$\lambda$1427 (P), SV$\lambda$1502 (P),
SiII$\lambda$1526 (I), CIV$\lambda$1550 (W plus I), CI$\lambda$1560
(I), FeII$\lambda$1608 (I), HeII$\lambda$1640 emission (W),
AlII$\lambda$1671 - possibly blended with CI$\lambda$1657 - (I),
NIV$\lambda$1720 (W), AlIII$\lambda$1859 (I plus W),
SiIII$\lambda$1892 (P), CIII]$\lambda$1909 (nebular emission-line),
FeIII$\lambda$1925 (P), and FeIII$\lambda$1960 (P). In the above,
the letters I, P, and W denote lines that are primarily of
interstellar, stellar photospheric, or stellar wind origin in
spectra of typical starbursts. The strong emission-line near
1200\AA\ is primarily due to geocoronal Ly$\alpha$.}

\figcaption{a) Plot of the log of starburst oxygen abundance 
by number (where
the solar value is 8.93) {\it vs.} the spectral slope in the UV
(where $F_{\lambda} \propto\ lambda^{\beta}$).
Since $\beta$ is primarily a reddening indicator, this figure
shows that the more metal-rich starbursts are more heavily
reddened. Unreddened starbursts should have $\beta \sim$ -2.0 to -2.6. 
b) Starburst metallicity {\it vs.} the ratio of the IR and
UV luminosities. Since this ratio indicates the amount of UV
extinction, more metal-rich starbursts are more severely dust-
extinguished. Note that in this and all subsequent figures, the solid dots
are the starbursts with the best IUE spectra ((S/N)$>$ 15), the
hollow dots are the starbursts with moderately good IUE spectra (S/N $= 10 -
 15$), and the dashed diagonal line is a least-squares fit to the data points.}

\figcaption{The mean equivalent width of the SiIV$\lambda$1400 and
CIV$\lambda$1550 lines (primarily due to stellar winds in most
starbursts) plotted {\it vs.} the starburst oxygen abundance. The
more metal-rich starbursts have stronger lines. See caption to Fig. 2
for symbol definition.}

\figcaption{The mean equivalent width of the SiII$\lambda$1260,
OI$\lambda$1302 plus SiII$\lambda$1304, and CII$\lambda$1335 lines
(primarily due to interstellar gas in most starbursts) plotted {\it
vs.} the starburst oxygen abundance. The more metal-rich starbursts
have stronger lines. See caption to Fig. 2 for symbol definition.}

\figcaption{a) The spectral slope in the UV ($\beta$) {\it vs.} the
mean equivalent width of the SiII$\lambda$1260,
OI$\lambda$1302 plus SiII$\lambda$1304, and CII$\lambda$1335 lines
(primarily interstellar in origin). The more heavily reddened
starbursts have stronger lines.
b) Same as a), but now plotted {\it
vs.} the mean equivalent width of the  SiIV$\lambda$1400 and
CIV$\lambda$1550 lines (primarily due to stellar winds).
Again, the more heavily
reddened starbursts have stronger absorption-lines.
See caption to Fig. 2 for symbol definition.}

\figcaption{a) The ratio of the IR and UV luminosities plotted {\it
vs.} the mean equivalent width of the SiII$\lambda$1260,
OI$\lambda$1302 plus SiII$\lambda$1304, and CII$\lambda$1335 lines
(primarily interstellar in origin). b) Same as a), but plotted {\it
vs.} the mean equivalent width of the SiIV$\lambda$1400 and
CIV$\lambda$1550 lines (primarily due to stellar winds). The more
heavily dust-extinguished starbursts have stronger lines in both
cases. See caption to Fig. 2 for symbol definition.}

\figcaption{The mean equivalent width of the SiIV$\lambda$1400 and
CIV$\lambda$1550 lines (primarily due to stellar winds) plotted
{\it vs.} the mean equivalent width of the SiII$\lambda$1260,
OI$\lambda$1302 plus SiII$\lambda$1304, and CII$\lambda$1335 lines
(primarily interstellar in origin). The strengths of the (largely)
stellar and (largely) interstellar absorption-lines are well-correlated.
See caption to Fig. 2 for symbol definition.}

\figcaption{The sum of the IR and UV luminosity (the {\it intrinsic}
UV luminosity of the starburst) plotted {\it vs.} the oxygen
abundance. The more metal-rich starbursts are more luminous.
See caption to Fig. 2 for symbol definition.}

\figcaption{a) The sum of the IR and UV luminosity (roughly, the {\it
intrinsic} UV luminosity of the starburst) plotted {\it vs.} the
UV spectral slope. b) As in a), but plotted {it vs.} the ratio of
the IR and UV luminosities. These figures show that the more
luminous starbursts are more heavily reddened and extinguished by
dust. See caption to Fig. 2 for symbol definition.}

\figcaption{a) The sum of the IR and UV luminosity of the starburst
plotted {\it vs.} the mean equivalent width of the
SiIV$\lambda$1400 and CIV$\lambda$1550 lines (primarily due to
stellar winds). b) Same as a), but plotted {\it vs.} the mean
equivalent width of the SiII$\lambda$1260, OI$\lambda$1302 plus
SiII$\lambda$1304, and CII$\lambda$1335 lines (primarily
interstellar in origin). In both cases, the more luminous
starbursts have stronger absorption-lines.
See caption to Fig. 2 for symbol definition.}

\figcaption{A montage showing the correlations between the
inclination-corrected HI$\lambda$21cm line width ($dV_{0} \simeq$
twice the rotation speed of the starburst `host' galaxy) and the
primary starburst parameters discussed in this paper. The more
massive (more rapidly-rotating) galaxies host starbursts that are
more luminous, more metal-rich, more heavily-reddened and dust-
extinguished, and have stronger UV absorption-lines (of both
interstellar and stellar wind origin). See caption to Fig. 2 for symbol 
definition.}

\figcaption{A montage showing the correlations between the absolute
blue magnitude of the starburst plus its host galaxy and the
primary starburst parameters discussed in this paper. The
optically-brighter galaxies host starbursts that are more luminous,
more metal-rich, more heavily-reddened and dust-extinguished, and have
stronger UV absorption-lines (of both interstellar and stellar wind
origin). See caption to Fig. 2 for symbol definition.}

\newpage
\parindent=0mm

{\tiny
\begin{deluxetable}{llrlrrclrr}
\tablewidth{0pt}
\tablenum{1}
\tablecaption{The {\it IUE} Galaxy Sample}
\tablehead{
\colhead{Galaxy} &
\colhead{Morphological} &
\colhead{$Z$} &
\colhead{Ref. $Z$} &
\colhead{$V_h$$\!\!\!\!\!\!\!\!$} &
\colhead{$D$$\!\!\!\!\!\!\!$} &
\colhead{$E(B$-$V)_{Gal}$} &
\colhead{$M_B$} &
\colhead{$dV_0$$\!\!\!$} &
\colhead{Log($L_{IR}$)$\!\!\!\!\!\!$} \\
\colhead{Name} &
\colhead{Type} &
\colhead{} &
\colhead{} &
\colhead{(km s$^{-1}$)$\!\!\!\!\!\!\!\!$} &
\colhead{(Mpc)$\!\!\!\!\!\!\!$} &
\colhead{(mag)} &
\colhead{(mag)} &
\colhead{(km)$\!\!\!$} &
\colhead{(erg s$^{-1}$)$\!\!\!\!\!\!$}}
\startdata

NGC1140     &IBm pec: Sy2    & 8.0 &G33               &1509 & 19.6 &0.03\pp &$-$
18.74  &275 & 42.87\nl
NGC1510     &SA0$^0$ pec?    & 7.9 &G34               & 989 & 11.3 &0.00\pp &$-$
16.79  &\nodata  & 41.81\nl
NGC1705     &SA0- pec:       & 8.0 &G34               & 597 &  5.7 &0.04\pp &$-$
16.19  &205 & 41.33\nl
UGC4483     &Im:             & 7.5 &32                & 156 &  3.5 &0.03\pp &$-$
12.90  & 70 &$\!\!\!\!<$39.66\nl
IZw18       &Compact         & 7.2 &11,31             & 756 & 14.3 &0.00\pp &$-$
14.52  & 79 &$\!\!\!\!<$41.89\nl
NGC3125     &S...            & 8.3 &G37,G23,G8        &1110 & 13.8 &0.06\pp &$-$
17.45  &\nodata  & 42.75\nl
UGC5720     &Im pec:         & 8.4 &22                &1461 & 24.5 &0.00\pp &$-$
18.54  &165 & 43.20\nl
NGC3353     &BCD/Irr         & 8.4 &G20               & 944 & 17.5 &0.00\pp &$-$
17.96  &135 & 42.96\nl
MRK153      &S?              & 7.8 &22                &2447 & 37.1 &0.00\pp &$-$
17.85  &\nodata  &$\!\!\!\!<$42.72\nl
NGC3738     &Im              & 8.4 &G18               & 229 &  4.4 &0.00\pp &$-$
16.09  &118 & 41.41\nl
NGC3991     &Sc              & 8.5 &G20               &3192 & 46.3 &0.00\pp &$-$
19.85  &163 & 43.57\nl
ESO572-34   &IBm             & 8.2 &23                &1075 & 13.4 &0.02\pp &$-$
16.56  &\nodata  & 41.97\nl
NGC4214     &IAB(s)m         & 8.2 &45                & 291 &  3.4 &0.00\pp &$-$
17.44  &178 & 42.05\nl
UGCA281     &S?              & 7.9 &40                & 281 &  4.5 &0.00\pp &$-$
13.13  &101 & \nodata \nl
NGC4449     &IBm             & 8.4 &G36,G25           & 207 &  3.1 &0.00\pp &$-$
17.47  &190 & 42.34\nl
NGC4670     &SB(s)0/a pec:   & 8.2 &44                &1069 & 14.6 &0.01\pp &$-$
17.77  &219 & 42.52\nl
NGC4861     &SB(s)m          & 8.0 &15                & 846 & 10.7 &0.02* &$-$17
.35  &110 & 42.09\nl
MRK66       &BCG             & 8.3 &16                &6525 & 91.0 &0.00\pp &$-$
19.80  &\nodata  &$\!\!\!\!<$43.45\nl
NGC5253     &Im pec:;HII     & 8.2 &8,41              & 404 &  4.1 &0.04\pp &$-$
17.37  & 95 & 42.44\nl
UGC9560     &Pec             & 8.2 &G25,G13,G27       &1213 & 21.5 &0.00\pp &$-$
16.84  & 71 & 42.36\nl
MK499       &Im:             & 8.4 &16                &7695 &106.4 &0.00\pp &$-$
19.84  &\nodata  & 44.00\nl
TOL1924-416 &Pec;HII         & 8.1 &G37,6             &2818 & 36.5 &0.07\pp &$-$
19.81  &\nodata  & 43.03\nl
NGC7673     &(R')SAc? pec    & 8.5 &G9                &3401 & 47.3 &0.03\pp &$-$
20.34  &309 & 43.84\nl
HARO 15     &(R)SB0$^0$ pec? & 8.6 &34                &6407 & 84.2 &0.02\pp &$-$
21.21  &262 & 43.78\nl
NGC1097     &SB(s)b Sy1      & 9.0 &34                &1275 & 15.6 &0.04* &$-$20
.89  &418 & 43.87\nl
NGC1672     &SB(s)b Sy2      & 9.1 &43                &1350 & 15.5 &0.00\pp &$-$
20.67  &429 & 43.76\nl
NGC2782     &SAB(rs)a pec    & 8.8 &G21,G7            &2562 & 37.6 &0.00\pp &$-$
20.58  &440 & 43.88\nl
NGC2903     &SAB(rs)bc       & 9.3 &43,30             & 556 &  7.0 &0.01\pp &$-$
19.60  &441 & 43.11\nl
NGC3049     &SB(rs)ab        & 9.1 &38,G24            &1494 & 22.6 &0.01\pp &$-$
18.77  &257 & 42.93\nl
NGC3256     &Pec;merger;HII  & 9.0 &G34,G39,G1        &2738 & 35.0 &0.13\pp &$-$
21.12  &\nodata  & 44.82\nl
NGC3310     &SAB(r)bc pec    & 9.0 &G28               & 980 & 17.9 &0.00\pp &$-$
20.11  &301 & 43.79\nl
NGC3351     &SB(r)b          & 9.3 &43,26             & 778 &  7.8 &0.01\pp &$-$
18.98  &384 & 42.87\nl
NGC3504     &(R)SAB(s)ab     & 9.1 &G17,G19,G5        &1539 & 25.4 &0.00\pp &$-$
20.38  &288 & 43.91\nl
NGC3690     &IBm pec         & 8.8 &G20,G13,G4,G14,G2 &3033 & 45.0 &0.00\pp &$-$
21.28  &530 & 45.06\nl
NGC4194     &IBm pec         &$\!\!\!\!>$8.6 &G20,G2,G42        &2506 & 38.2 &0.
00\pp &$-$19.90  &118 & 44.27\nl
NGC4258     &SAB(s)bc        & 9.0 &43                & 448 &  6.8 &0.00\pp &$-$
20.05  &239 & 42.86\nl
NGC4321     &SAB(s)bc        & 9.3 &43,30             &1586 & 15.9 &0.01\pp &$-$
21.01  &552 & 43.60\nl
NGC4385     &SB(rs)0$^+$:    &$\!\!\!\!>$8.7 &G12,G35           &2140 & 31.8 &0.
01\pp &$-$19.34  &119 & 43.44\nl
NGC5236     &SAB(s)c         & 9.3 &43,10             & 516 &  3.8 &0.03\pp &$-$
19.82  &392 & 43.01\nl
NGC5996     &SB?             & 8.9 &G20               &3304 & 47.4 &0.01\pp &$-$
20.84  &211 & 43.79\nl
NGC6217     &(R)SB(rs)bc     & 9.0 &G3                &1362 & 23.6 &0.03\pp &$-$
20.22  &405 & 43.60\nl
NGC7250     &Sdm?            & 8.7 &34                &1166 & 19.5 &0.15\pp &$-$
18.84  &208 & 42.86\nl
NGC7496     &SB(s)b Sy2      & 9.0 &G34               &1649 & 20.5 &0.04* &$-$19
.81  &288 & 43.37\nl
NGC7552     &(R')SB(s)ab     & 9.2 &G34               &1585 & 19.6 &0.04* &$-$20
.38  &275 & 44.22\nl
NGC7714     &SB(s)b: pec     &$\!\!\!\!>$8.7 &G34,G20,G13,G29   &2798 & 37.9 &0.
04\pp &$-$20.04  &241 &43.93 \nl
\noalign{\vskip -1.0em}
\tablebreak

\tablenotetext{}{
References for $Z$ -- (G) Garnett's estimates based on published
data. Uses $T_e$ when available, if not, uses the Edmunds \& Pagel (1984)
calibration of [OII]+[OIII]/$H_\beta$ and [OIII]/[NII] for general purposes.
(1) Aguero \& Lipari (1991); (2) Armus, Heckman, \& Miley (1989);
(3) Ashby, Houck, \& Hacking (1992); (4) Augarde \& Lequeux (1985);
(5) Balzano (1983); (6) Bergvall (1985); (7) Boer, Schulz, \& Keel (1992);
(8) Campbell, Terlevich, \& Melnick (1986);
(9) Duflot-Augarde \& Alloin (1982); (10) Dufour \etal (1980);
(11) Dufour, Garnett, \& Shields (1988); (12) Durret \& Tarrab (1988);
(13) French (1980; Note that abundances from French are based
on temperature measured from [OII] $\lambda$3727/7324. The upper levels of
OII can be populated significantly by dielectronic recombinations, which can
give enhanced [OII] $\lambda$7324.); (14) Friedman \etal (1987); (15) 
Dinerstein \& Shields (1986); Garnett
(1990); (16) Hartmann \etal (1988; Note that Hartmann \etal use only
[0 III]/$H_\beta$, which is not very reliable for individual HII regions,
because [O III]/$H_\beta$ is rather dependent on HII region structural
parameters as well as abundance. It also uses an old calibration.);
(17) Heckman \etal (1983); (18) Hunter, Gallagher, \& Rautenkranz (1982);
(19) Keel (1984); (20) Kennicutt (1992); (21) Kinney \etal (1984);
(22) Kunth \& Joubert (1985); (23) Kunth \& Sargent (1983);
(24) Kunth \& Schild (1986); (25) Lequeux \etal (1979); (26) McCall, Rybski,
\& Shields (1985); (27) O'Connell, Thuan, \& Goldstein (1978);
(28) Pastoriza \etal (1993); (29) de Robertis \& Shaw (1988);
(30) Shields, Kennicutt, \& Skillman (1991); (31) Skillman \& Kennicutt (1993);
(32) Skillman \etal (1994); (33) Stasinska, Comte, \& Vigroux (1986);
(34) Storchi-Bergmann, Kinney, \& Challis (1995); (35) Sugai \& Taniguchi
(1992);
(36) Talent (1980); (37) Terlevich \etal (1991); (38) Vacca \& Conti (1992);
(39) V\'eron-Cetty \& V\'eron (1986); (40) Viallefond \& Thuan (1983);
(41) Walsh \& Roy (1989); (42) Zamorano \& Rego (1985);
(43) Zaritzky, Kennicutt, \& Huchra (1994); (44) This study; (45) 
Kobulnicky \& Skillman (1996).}

\enddata
\end{deluxetable}
\newpage
\begin{deluxetable}{lcccccrr}
\tablewidth{0pt}
\tablenum{2}
\tablecaption{Ultraviolet-Related Properties}
\tablehead{
\colhead{Galaxy} &
\colhead{$\beta$} &
\colhead{Log($L_{UV}$)} &
\colhead{Log($L_{IR}/L_{UV}$)} &
\colhead{Log($L_{UV}+L_{IR}$)} &
\colhead{$W_{IS}$} &
\colhead{$W_W$$\!\!\!\!$} \\
\colhead{Name} &
\colhead{} &
\colhead{(erg s$^{-1}$)} &
\colhead{} &
\colhead{(erg s$^{-1}$)} &
\colhead{(\AA)} &
\colhead{(\AA)$\!\!\!\!\!\!\!\!\!\!\!$}} 
\startdata
 
NGC1140     &$-$1.54 &42.47 &$\phantom{<\!\!-}$0.40 &$\phantom{<}$43.02    &2.7
& 3.4 \nl
NGC1510     &$-$1.48 &41.66 &$\phantom{<\!\!-}$0.15 &$\phantom{<}$42.04    &3.5
& 2.9 \nl
NGC1705     &$-$2.64 &41.87 &$\phantom{<}$$-$0.54 &$\phantom{<}$41.98    &2.4 &
4.1  \nl
UGC4483     &$-$2.18 &40.09 &$\phantom{-}$$<$0.57 &$<$40.77 &3.0 & 1.6 
\nl
IZw18       &$-$2.07 &41.61 &$\phantom{-}$$<$0.28 &$<$42.07 &2.9 & 0.8
\nl
NGC3125     &$-$0.72 &42.02 &$\phantom{<\!\!-}$0.73  &$\phantom{<}$42.82    &4.0
 & 2.9  \nl
UGC5720     &$-$1.25 &42.61 &$\phantom{<\!\!-}$0.60  &$\phantom{<}$43.30    &3.9
 & 5.5 \nl
NGC3353     &$-$1.26 &42.23 &$\phantom{<\!\!-}$0.73  &$\phantom{<}$43.04    &3.2
 & 2.8 \nl
MRK153      &$-$2.31 &42.76 &$<\!\!-$0.04 &$<$43.04 &2.9 &4.3 \nl
NGC3738     &$-$1.60 &41.02 &$\phantom{<\!\!-}$0.39  &$\phantom{<}$41.55    &2.6
 & 4.0 \nl
NGC3991     &$-$1.89 &43.18 &$\phantom{<\!\!-}$0.39  &$\phantom{<}$43.72    &1.9
 & 3.3 \nl
ESO572-34   &$-$2.22 &41.71 &$\phantom{<\!\!-}$0.26  &$\phantom{<}$42.16    &3.0
 & 4.7 \nl
NGC4214     &$-$1.53 &41.23 &$\phantom{<\!\!-}$0.82  &$\phantom{<}$42.11    &2.4
 & 3.5 \nl
UGCA281     &$-$1.87 &40.83 &$\phantom{-}$\nodata &$\phantom{<}$\nodata  &2.4 &
1.3 \nl
NGC4449     &$-$2.14 &41.19 &$\phantom{<\!\!-}$1.15  &$\phantom{<}$42.37    &2.3
 & 3.7 \nl
NGC4670     &$-$1.41 &42.31 &$\phantom{<\!\!-}$0.20  &$\phantom{<}$42.73    &3.4
 & 4.7 \nl
NGC4861     &$-$2.48 &42.20 &$\phantom{<}$$-$0.11 &$\phantom{<}$42.45    &1.6 &
2.9 \nl
MRK66       &$-$1.67 &43.07 &$\phantom{-}$$<$0.38 &$<$43.60 &2.6 & 4.5
\nl
NGC5253     &$-$1.13 &41.68 &$\phantom{<\!\!-}$0.77  &$\phantom{<}$42.51    &3.1
 & 3.5 \nl
UGC9560     &$-$2.06 &42.27 &$\phantom{<\!\!-}$0.09  &$\phantom{<}$42.62    &2.3
 & 3.4 \nl
MK499       &$-$1.07 &43.27 &$\phantom{<\!\!-}$0.73  &$\phantom{<}$44.07    &4.2
 & 5.3 \nl
TOL1924-416 &$-$2.06 &42.95 &$\phantom{<\!\!-}$0.08  &$\phantom{<}$43.30    &1.5
 & 3.0 \nl
NGC7673     &$-$1.14 &42.94 &$\phantom{<\!\!-}$0.89  &$\phantom{<}$43.89    &2.0
 & 4.0 \nl
HARO 15     &$-$1.28 &43.36 &$\phantom{<\!\!-}$0.42  &$\phantom{<}$43.92    &2.3
 & 5.7 \nl
NGC1097     &$-$0.31 &42.09 &$\phantom{<\!\!-}$1.78  &$\phantom{<}$43.88    &3.7
 & 7.9 \nl
NGC1672     &$-$0.06 &42.18 &$\phantom{<\!\!-}$1.58  &$\phantom{<}$43.77    &4.9
 & 8.3  \nl
NGC2782     &$-$0.99 &42.68 &$\phantom{<\!\!-}$1.20  &$\phantom{<}$43.90    &4.5
 & 5.3 \nl
NGC2903     &$\phantom{-}$0.84 &41.33 &$\phantom{<\!\!-}$1.78  &$\phantom{<}$43.
12    &6.6 & 7.9 & \nl
NGC3049     &$-$0.98 &42.08 &$\phantom{<\!\!-}$0.85  &$\phantom{<}$42.99    &3.0
 & 8.5 \nl
NGC3256     &$\phantom{-}$0.44 &42.78 &$\phantom{<\!\!-}$2.05  &$\phantom{<}$44.
83    &5.6 & 8.5 \nl
NGC3310     &$-$0.72 &42.73 &$\phantom{<\!\!-}$1.06  &$\phantom{<}$43.83    &5.2
 & 5.7 & \nl
NGC3351     &$\phantom{-}$0.13 &41.47 &$\phantom{<\!\!-}$1.39  &$\phantom{<}$42.
88    &4.1 & 9.1 \nl
NGC3504     &$-$0.62 &42.51 &$\phantom{<\!\!-}$1.40  &$\phantom{<}$43.92    &3.7
 & 5.3 \nl
NGC3690     &$-$0.82 &42.89 &$\phantom{<\!\!-}$2.17  &$\phantom{<}$45.06    &4.2
 & 4.8 \nl
NGC4194     &$-$0.02 &42.68 &$\phantom{<\!\!-}$1.59  &$\phantom{<}$44.28    &6.3
 & 7.4 \nl
NGC4258     &$-$0.81 &41.27 &$\phantom{<\!\!-}$1.60  &$\phantom{<}$42.88    &4.8
 & 4.5 \nl
NGC4321     &$-$0.53 &42.00 &$\phantom{<\!\!-}$1.60  &$\phantom{<}$43.61    &4.8
 & 9.8 \nl
NGC4385     &$-$0.46 &42.44 &$\phantom{<\!\!-}$1.00  &$\phantom{<}$43.48    &4.4
 & 7.8 \nl
NGC5236     &$-$0.76 &41.80 &$\phantom{<\!\!-}$1.21  &$\phantom{<}$43.03    &5.2
 & 8.9 \nl
NGC5996     &$-$0.73 &42.73 &$\phantom{<\!\!-}$1.05  &$\phantom{<}$43.82    &3.9
 & 8.6 \nl
NGC6217     &$-$0.60 &42.35 &$\phantom{<\!\!-}$1.25  &$\phantom{<}$43.63    &5.2
 &10.1 \nl
NGC7250     &$-$1.20 &42.13 &$\phantom{<\!\!-}$0.73  &$\phantom{<}$42.94    &4.5
 & 4.3 \nl
NGC7496     &$-$1.18 &42.08 &$\phantom{<\!\!-}$1.29  &$\phantom{<}$43.39    &4.0
 & 7.1 \nl
NGC7552     &$\phantom{-}$0.48 &42.29 &$\phantom{<\!\!-}$1.93  &$\phantom{<}$44.
23    &6.7 & 9.0 \nl
NGC7714     &$-$1.03 &42.99 &$\phantom{<\!\!-}$0.95  &$\phantom{<}$43.98    &3.7
 & 6.3

\enddata
\end{deluxetable}
}
\newpage
\begin{deluxetable}{lcccccccc}
\tablewidth{0pt}
\tablenum{3}
\tablecaption{Summary of Correlation Coefficients}
\tablehead{
\colhead{} &
\colhead{$\beta$} &
\colhead{$Z$} &
\colhead{Log($L_{IR}/L_{UV}$)} &
\colhead{Log($L_{UV}+L_{UV}$)} &
\colhead{$dV_{0}$} &
\colhead{$W_{IS}$} &
\colhead{$W_{W}$} &
\colhead{$M_{B}$} }
\startdata
 
$\beta$        &\nodata &0.80 &0.85 &0.60 &0.59 &0.84 &0.76 &0.63 \nl
$Z$            &0.80 &\nodata &0.78 &0.59 &0.72 &0.66 &0.85 &0.77 \nl
Log$(L_{IR}/L_{UV}$) &0.85 &0.78 &\nodata &0.62 &0.66 &0.75 &0.66 &0.65 \nl
Log$(L_{UV}+L_{IR}$) &0.60 &0.59 &0.62 &\nodata &0.57 &0.46 &0.54 &0.91 \nl
$dV_{0}$       &0.59 &0.72 &0.66 &0.57 &\nodata &0.49 &0.64 &0.70 \nl
$W_{IS}$       &0.84 &0.66 &0.75 &0.46 &0.49 &\nodata &0.69 &0.47 \nl
$W_{W}$        &0.76 &0.85 &0.66 &0.54 &0.64 &0.69 &\nodata &0.69 \nl
$M_{B}$        &0.63 &0.77 &0.65 &0.91 &0.70 &0.47 &0.69 &\nodata

\enddata
\end{deluxetable}

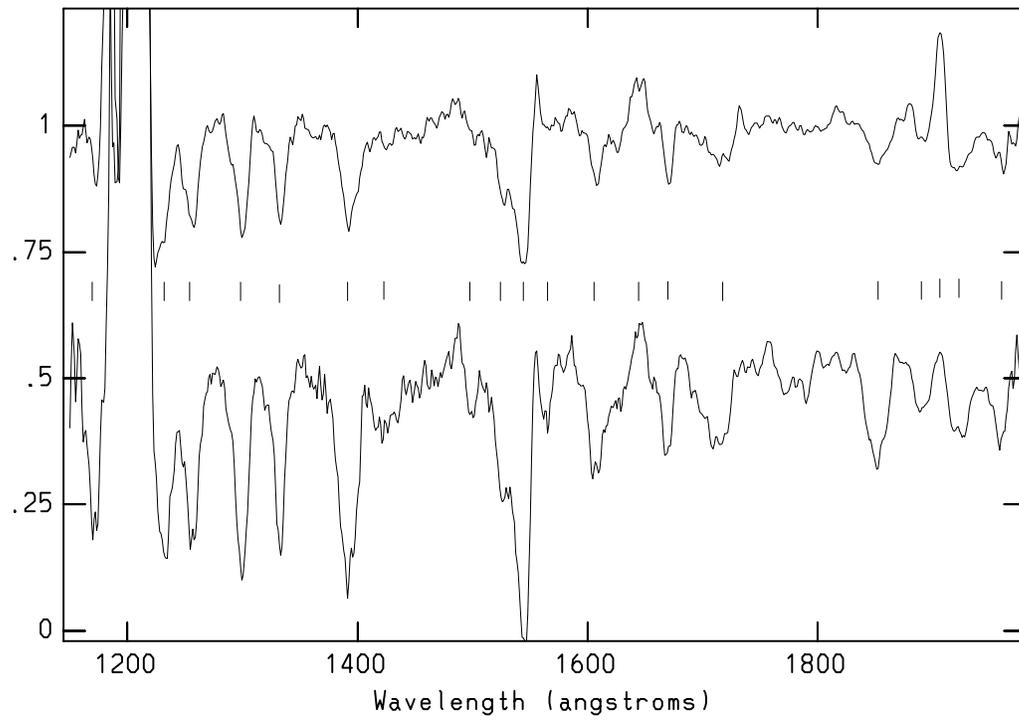
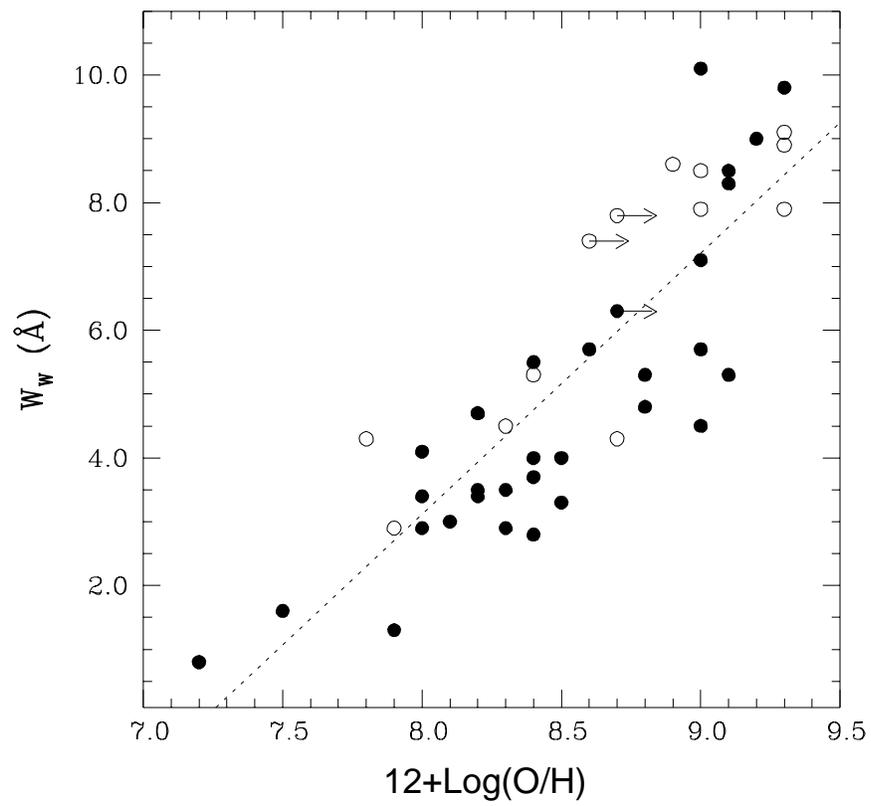
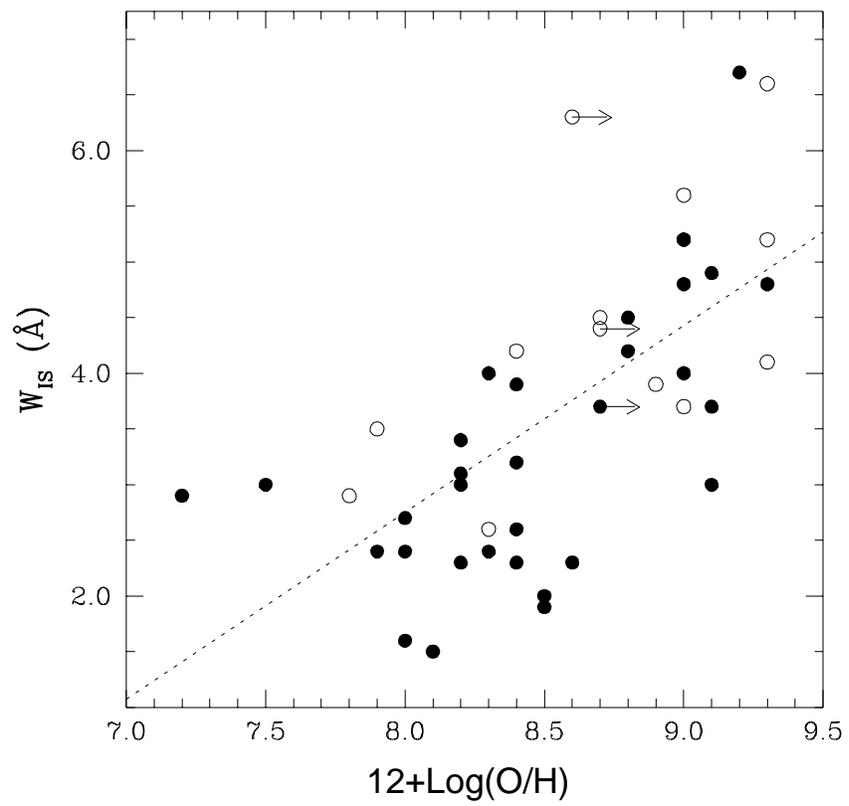
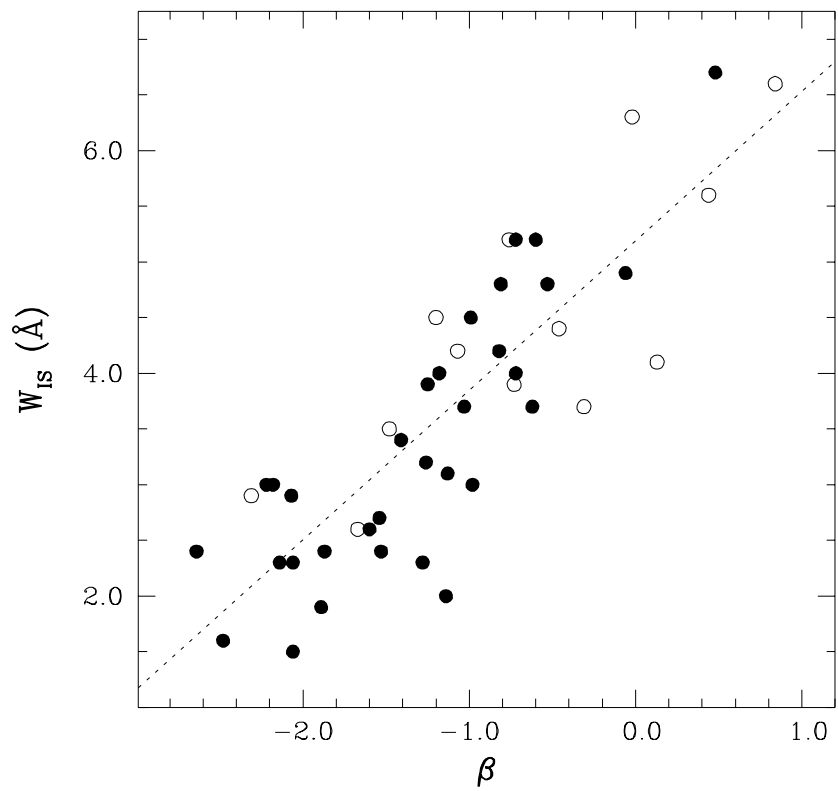
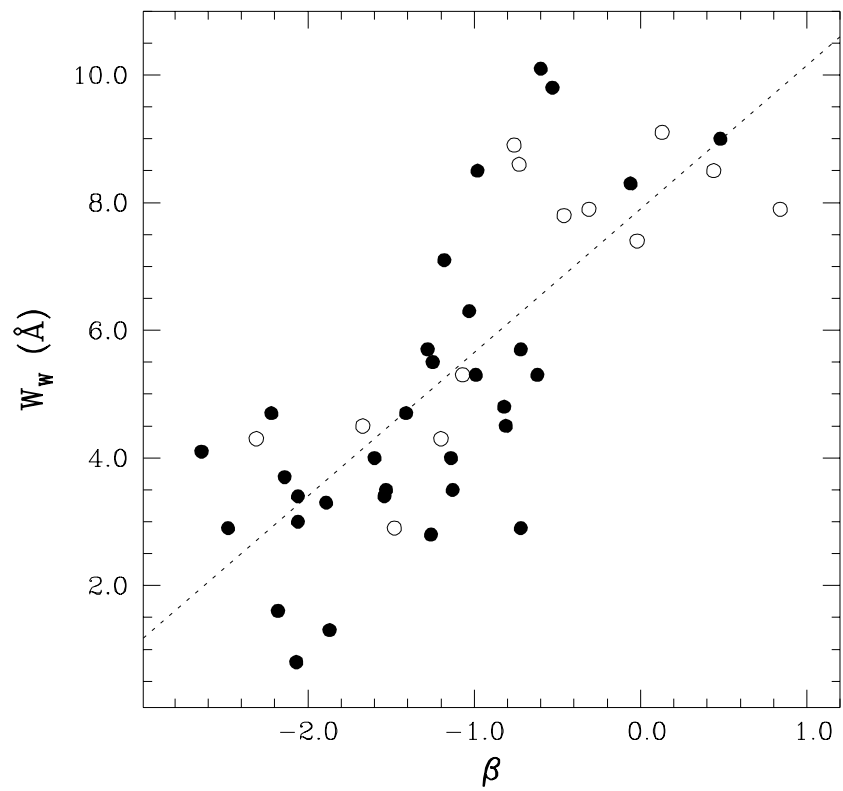
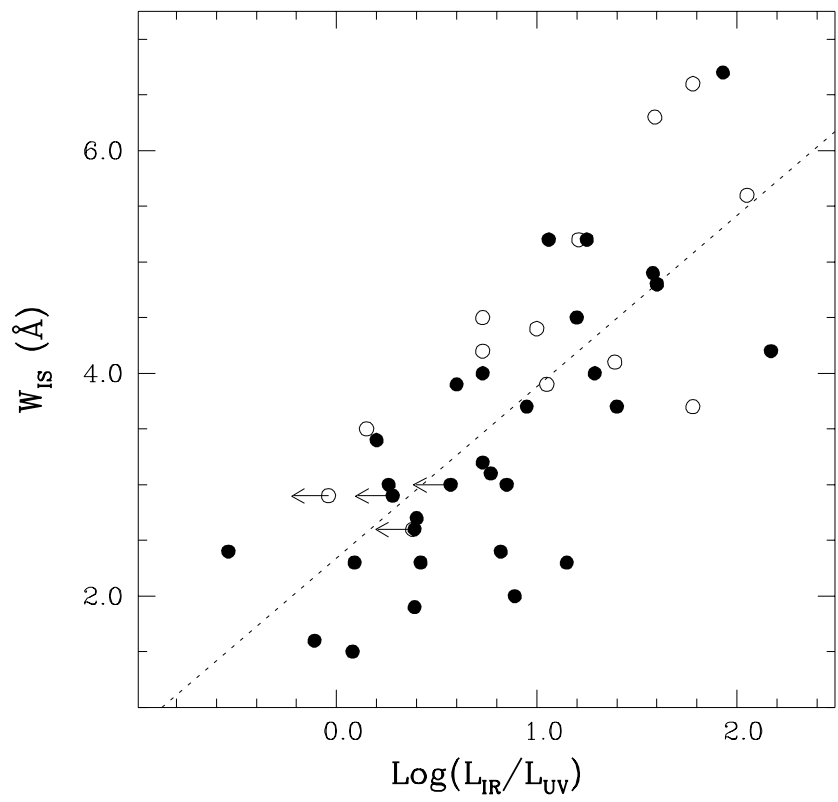
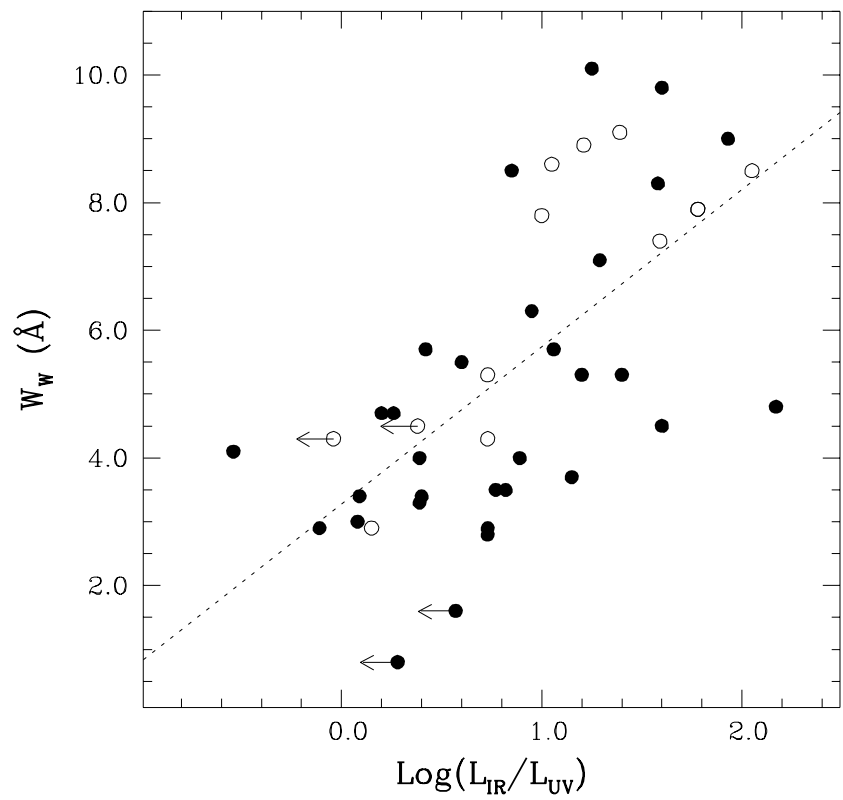
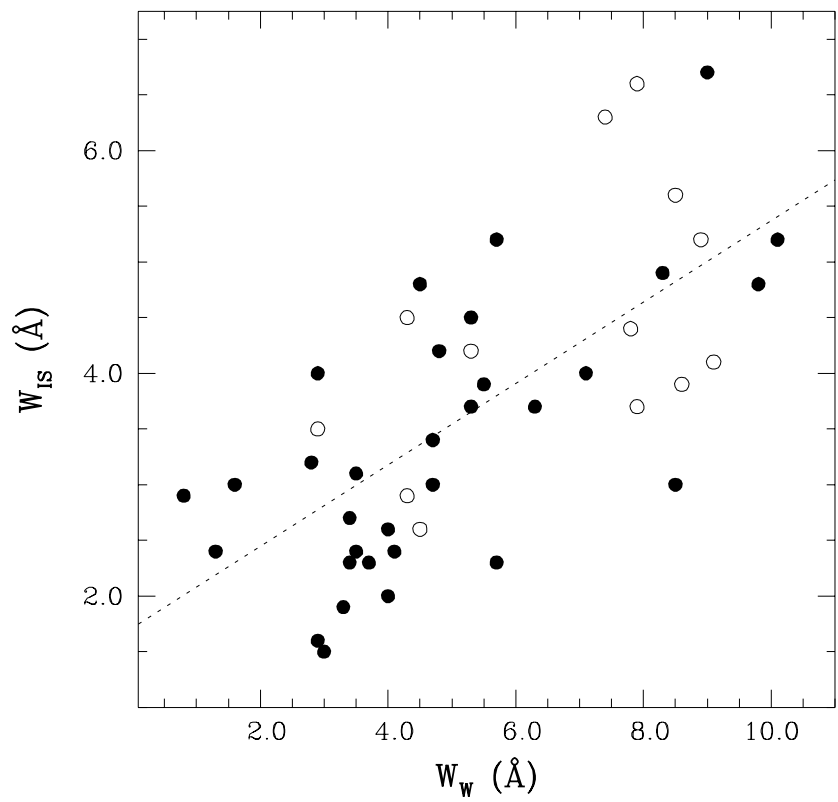
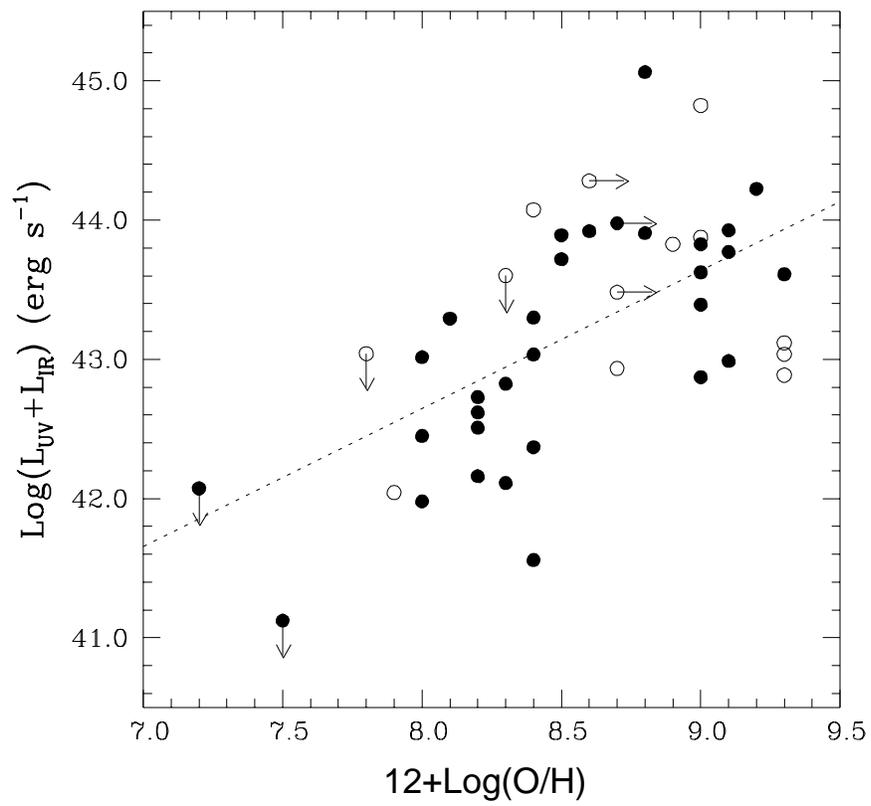
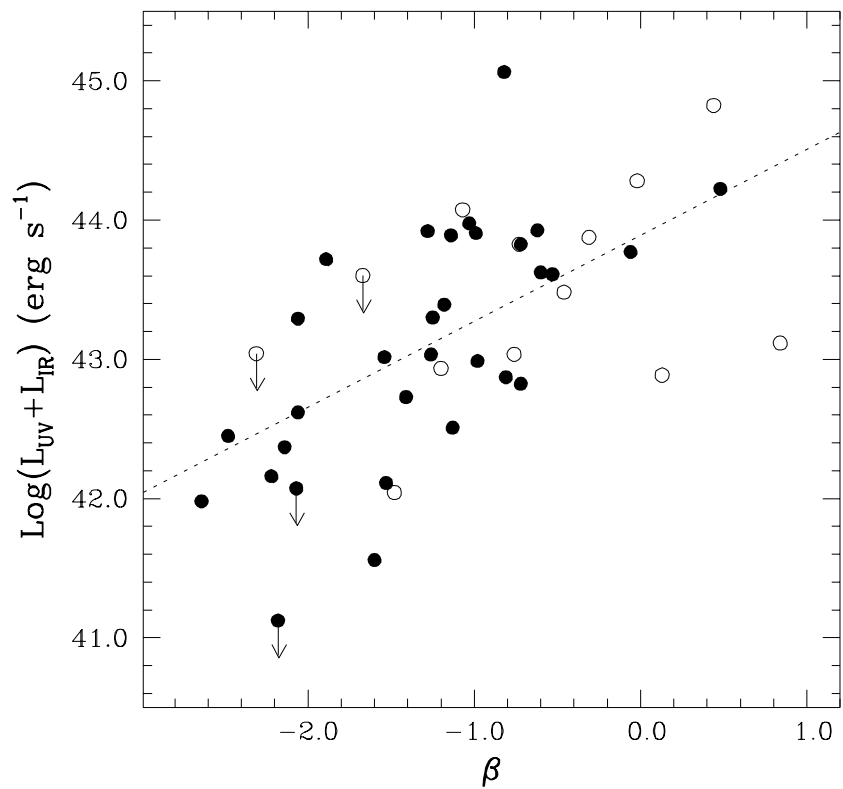
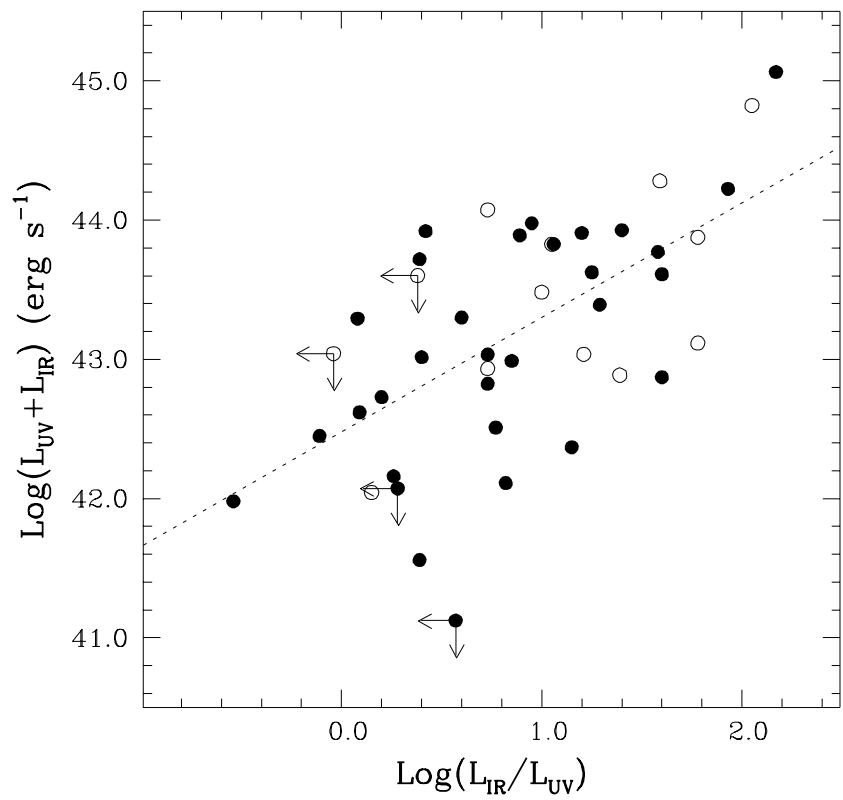
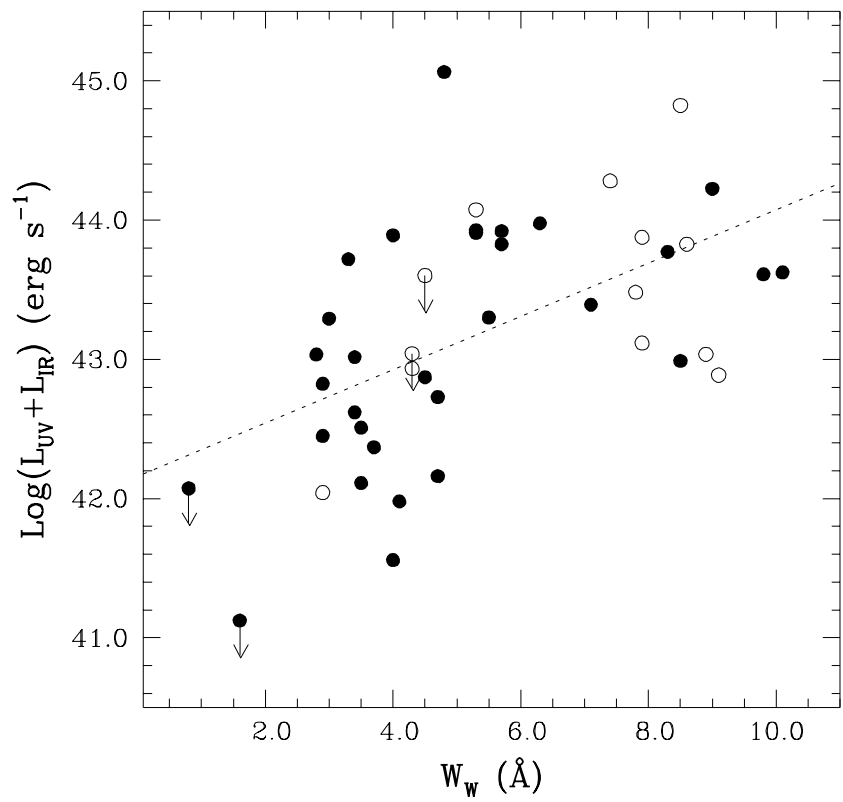
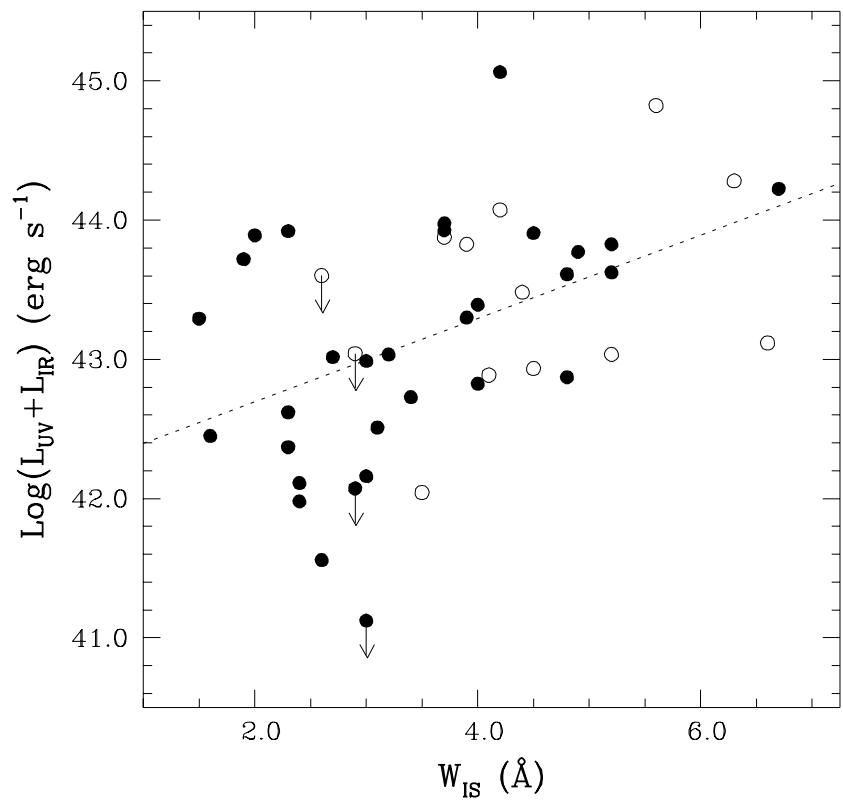
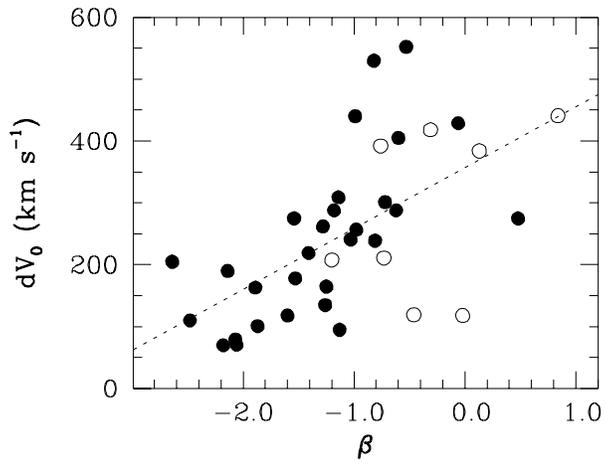
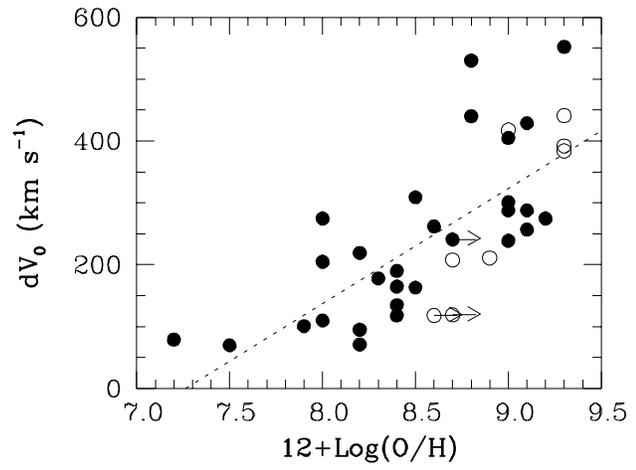
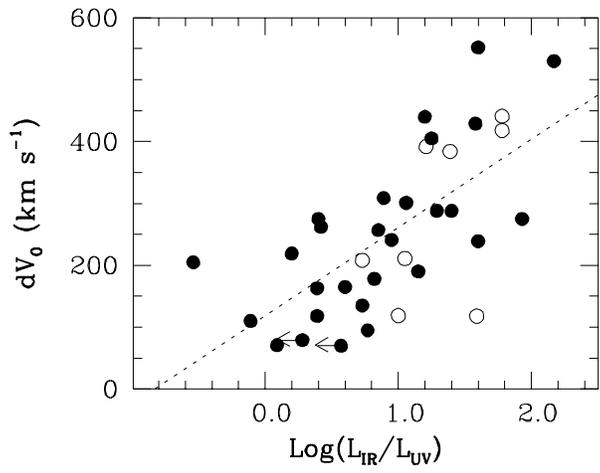
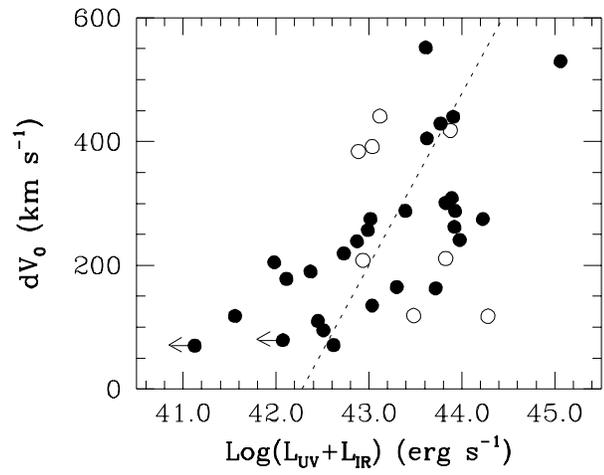
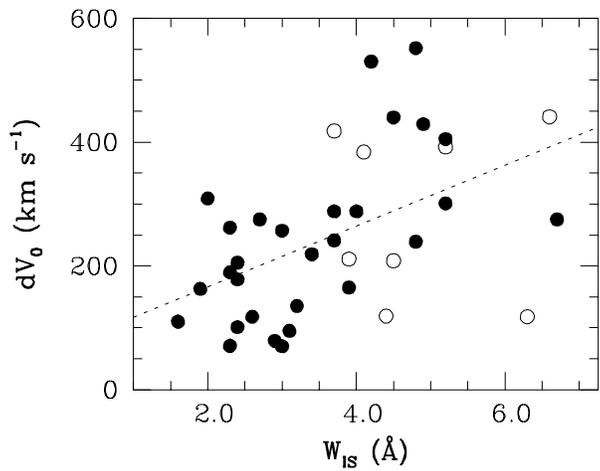
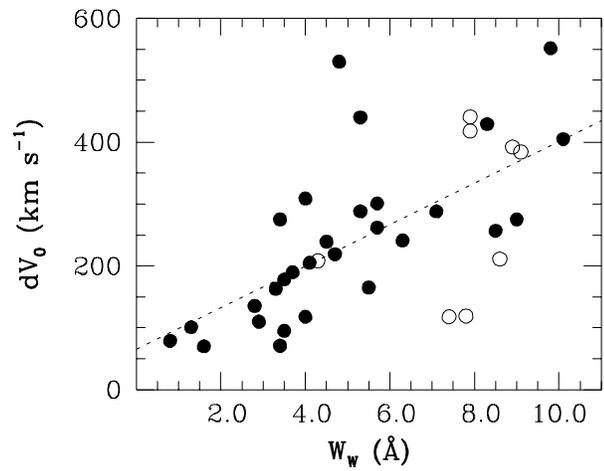
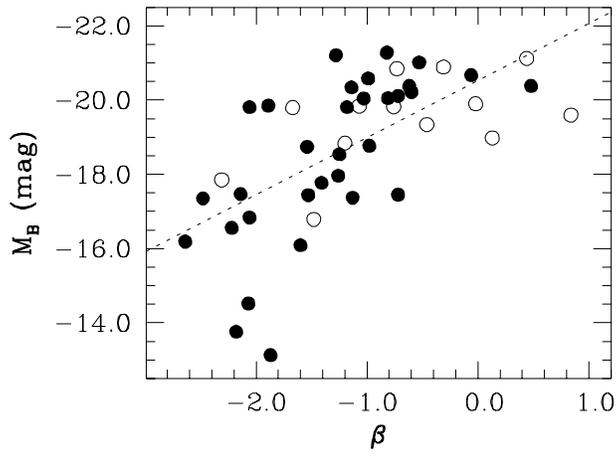
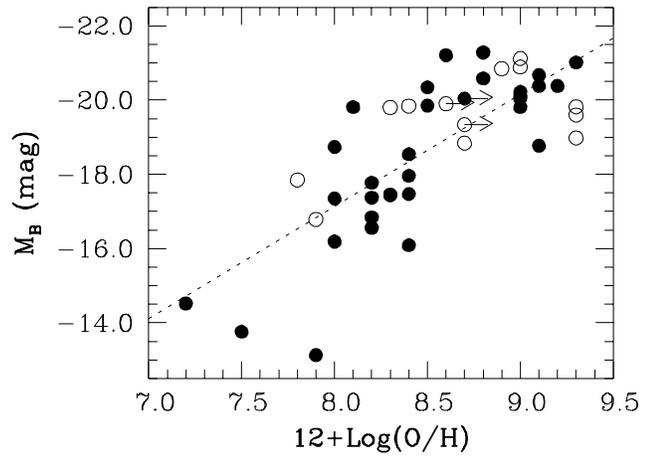
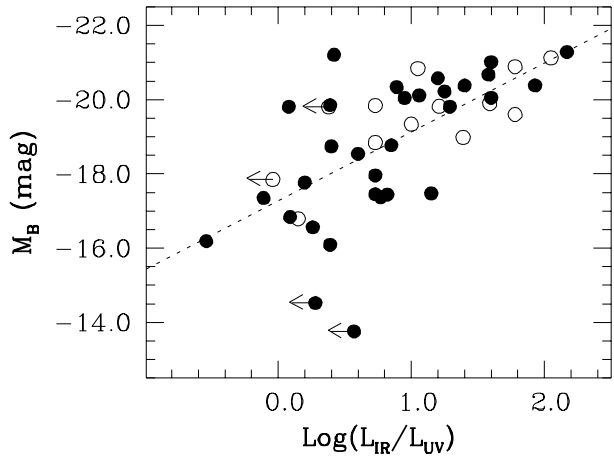
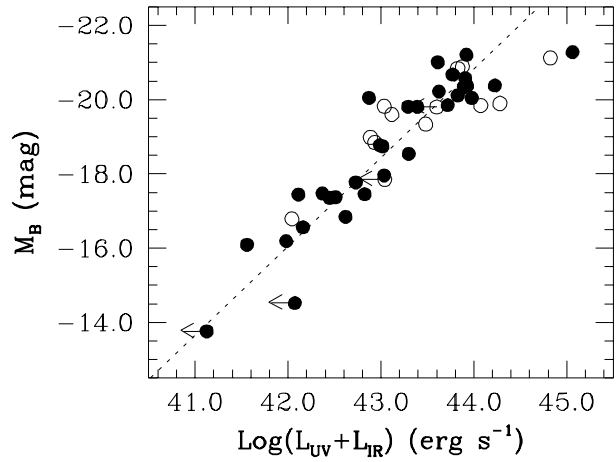
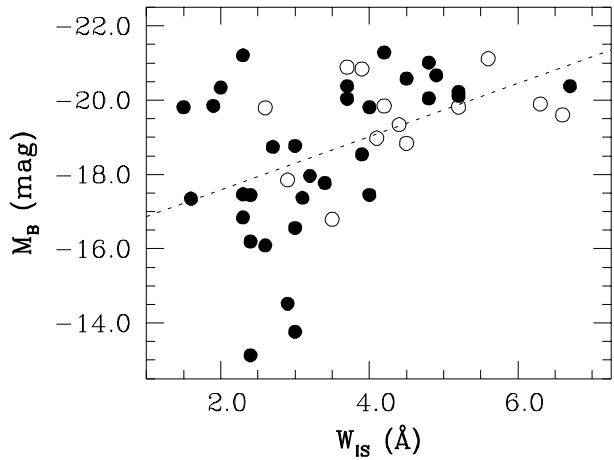
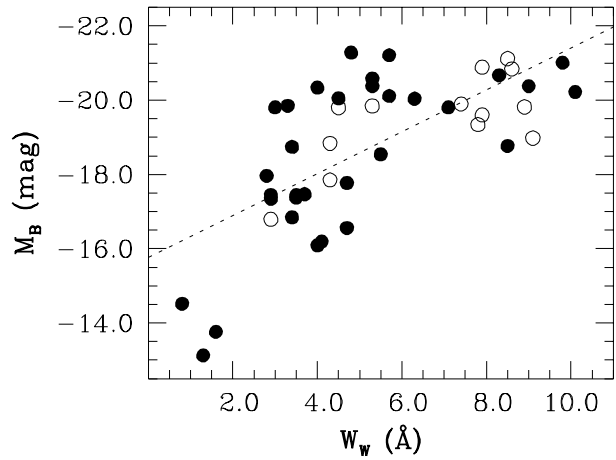
\end{document}